\documentclass{article}

\usepackage{arxiv}

\usepackage[utf8]{inputenc} % allow utf-8 input
\usepackage[T1]{fontenc}    % use 8-bit T1 fonts
\usepackage{hyperref}       % hyperlinks
\usepackage{url}            % simple URL typesetting
\usepackage{amsfonts}       % blackboard math symbols
\usepackage{graphicx} % Required for inserting images
\usepackage{nicefrac}       % compact symbols for 1/2, etc.
\usepackage{amsmath}
\usepackage{amssymb}
\usepackage{xcolor}
\usepackage{subfig}
\usepackage{natbib}
\usepackage{doi}
\usepackage{bm}     % bold math symbols
\usepackage{tikz}   % conceptual diagram
\usetikzlibrary{shapes.geometric}  % ellipses

\newcommand{\bnu}{\boldsymbol{\nu}}

\newcommand{\btheta}{\boldsymbol\theta}
\newcommand{\numberthis}{\addtocounter{equation}{1}\tag{\theequation}}

\usepackage{ulem}  %for strikethough

\title{A Proxy-likelihood Estimator for Multivariate Extremes Models with Intractable Likelihoods}
\author{
    Troy P. Wixson \\
        Department of Mathematics and Statistics\\
        University of Massachusetts Amherst\\
        Amherst, Massachusetts 
        \And 
    Daniel Cooley \\
        Department of Statistics \\
        Colorado State University \\
        Fort Collins, Colorado}

\begin{document}
\maketitle

\begin{abstract}
Many multivariate extremes models have intractable likelihoods requiring practitioners to use alternative fitting methods. 
The tail pairwise dependence is a summary measure of the dependence in the tail of any multivariate regular variation model.
We develop an objective function for model fitting that relies on the tail pairwise dependence as the link between our desired model (that does not have a likelihood) and a proxy model (that has a likelihood).
We employ the bivariate H\"usler-Reiss distribution as the proxy model and show that there is a one-to-one relationship between the dependence parameter and the tail pairwise dependence value. 
Our proxy-likelihood estimator is fully developed for the transformed linear extremes time series (TLETS) models of \citet{mhatre_cooley2024} and is applied to the wildfire weather data of \citet{wixson_cooley2023attribution}.
Simulations demonstrate that the proxy-likelihood is a competitive TPD estimator, is better at fitting TLETS models than existing methods, and is amenable to likelihood-based model selection techniques.  
Our estimator has smaller bias when tail dependence is weak than existing estimators reducing the need for bias adjustments. 
Without these adjustments, we note an increase in the tail dependence in weather-related wildfire risk between past and present climates. 
\end{abstract}

\section{Introduction}

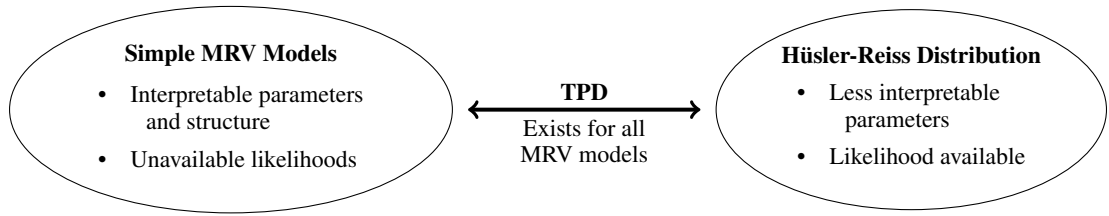
\begin{figure}[b]
    \centering
    {\small
    \begin{tikzpicture}
      % Left oval
      \node[draw, ellipse, minimum width=4cm, minimum height=2.5cm, align=center] (left) at (0,0) {
        \textbf{Simple MRV Models} \\[0.2cm]
        \begin{tabular}{l}
        \textbullet \quad Interpretable parameters \\
        \quad\quad and structure \\[0.15cm]
        \textbullet \quad Unavailable likelihoods
        \end{tabular}
      };
      
      % Right oval
      \node[draw, ellipse, minimum width=4cm, minimum height=2.5cm, align=center] (right) at (9,0) {
        \textbf{Hüsler-Reiss Distribution} \\[0.2cm]
        \begin{tabular}{l}
        \textbullet \quad Less interpretable \\
        \quad\quad parameters \\[0.15cm]
        \textbullet \quad Likelihood available
        \end{tabular}
      };
      
      % Double-headed arrow
      \draw[<->, line width=1.5pt, shorten >=0.2cm, shorten <=0.2cm] (left.east) -- (right.west) 
        node[midway, above, font=\bfseries] {TPD}
        node[midway, below, align=center] {Exists for all \\ MRV models};
  \end{tikzpicture}
  }
  \caption{Conceptual diagram, we link the MRV model parameters to the HR with the TPD.}
  \label{fig:frameworkDiagram}
\end{figure}

% \DC{Likelihood methods and extremes}
Many convenient and interpretable multivariate extremes models are not amenable to likelihood-based inference.
It is well known that the likelihood of multivariate max-stable models rapidly becomes intractable as the dimension grows, since the number of terms in the distribution's exponent measure increases as the Bell number \citep{castruccio2016high}.
Whereas the max-stable distributions are natural for modeling data which are componentwise block maxima, multivariate regular variation (MRV) is often used as a model for multivariate threshold exceedance data.
Several MRV models which have interpretable parameters and model structure are also difficult to reconcile with likelihood-based inference, albeit for different reasons.
In this work, we propose a method for fitting MRV models via a method summarized by Figure \ref{fig:frameworkDiagram}.
We use a tail pairwise dependence (TPD) measure to link MRV models' parameters to a H\"usler-Reiss (HR) Distribution, which has available bivariate likelihood expressions.

% \DC{MRV overview.}
MRV is fundamentally connected to classical extremes as it represents the domain of attraction of the multivarate max-stable distributions with heavy tailed margins.
From a modeling point-of-view, MRV is sensible for extremes as the framework only describes the joint tail.
Details of MRV will be given in Section 2.
For now, consider the scatterplots in Figures 1 and 2 which nicely illustrate some fundamental behavior of bivariate regular variation.
A multivariate regularly varying random vector is heavy tailed; one can see that all of the scatterplots have the bulk of their points near the origin, and the viewer's eye is naturally drawn to the behavior of the large points far from the origin.
One further notices that the distribution of the large points appears to be polar in nature; Section 2's definition will include an "angular measure" which describes in which directions large points are more (or less) likely to occur.
In two dimensions, the polar nature is easy to visualize, and the angular measure is relatively easy to model, but modeling and estimation become more difficult in higher dimensions.

\begin{figure}
    \centering
    \subfloat[$k = 5$]{\label{fig:max-lin5}\includegraphics[width=0.3\textwidth]{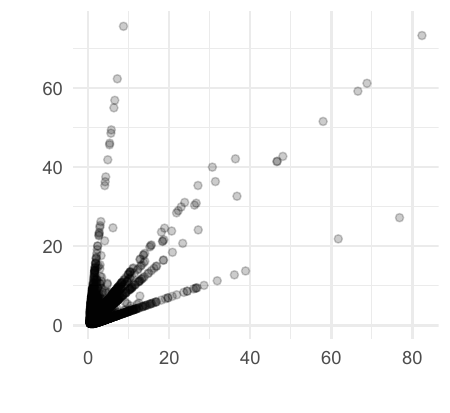}}
    \subfloat[$k = 15$]{\label{fig:max-lin15}\includegraphics[width=0.3\textwidth]{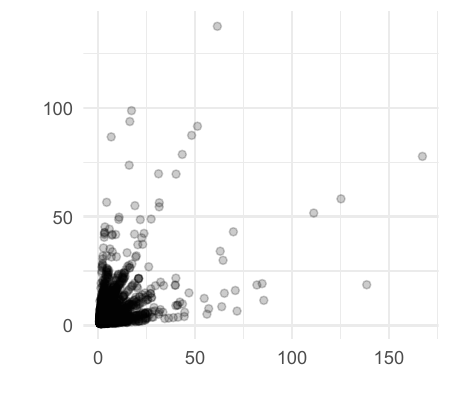}}
    \subfloat[$k = 25$]{\label{fig:max-lin25}\includegraphics[width=0.3\textwidth]{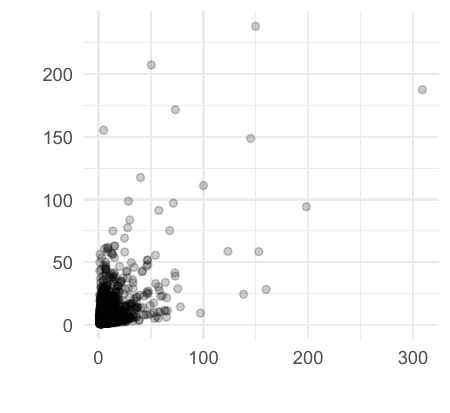}}
    \caption{Max-linear models have discrete angular measures. Point clouds were generated with 10000 points from max-linear modes with $k = 5, 15, 25$ point masses respectively. 
    }
    \label{fig:max-lin_pointclouds}
\end{figure}

\begin{figure}
    \centering
    	\subfloat[ERA5 lag-$1$ FWI]{\label{fig:late_ts_lag1}\includegraphics[width=0.3\textwidth]{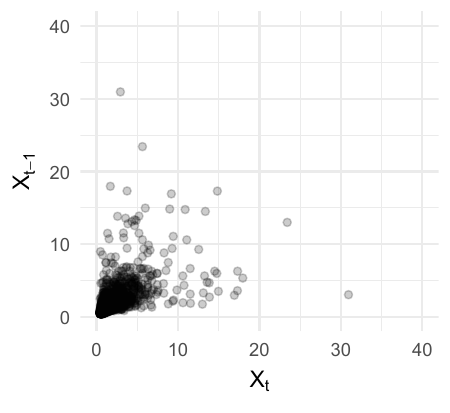}}
        \subfloat[ARMA($1,1$) generated lag-$1$]{\label{fig:arma11_lag1}\includegraphics[width=0.3\textwidth]{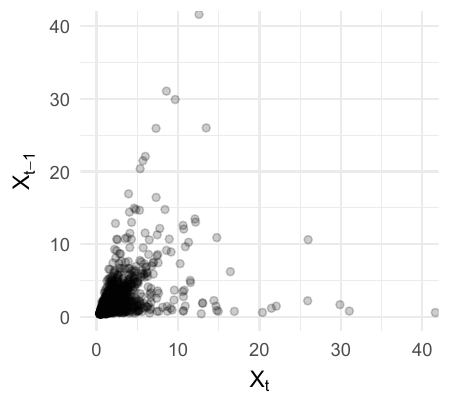}}
    	\subfloat[HR generated points]{\label{fig:hr_lag1_points}\includegraphics[width=0.3\textwidth]{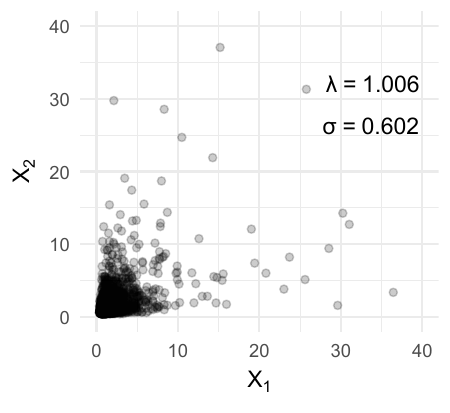}}
    \caption{Scatterplots have different angular measures but the same estimated/model TPD. (\ref{fig:late_ts_lag1}) appears to have a continuous angular measure like (\ref{fig:hr_lag1_points}). }
    \label{fig:angular_measure_mismatch}
\end{figure}

% \DC{Linear models}
For illustration, consider a linear or max-linear construction of MRV. 
Let $\bm Z = (Z_1, \ldots, Z_q)^\top$, where $Z_k, k = 1, \ldots, q$ are iid, non-negative, and regularly varying. 
For non-negative $A \in \mathbb{R}^{p \times q}$ with $k$th column $\bm a_{\cdot k} \neq \bm 0$ for all $k = 1, \ldots, q$, the linear construction $A \bm Z = \sum_{k = 1}^q \bm a_{\cdot k} Z_k$ is multivariate regularly varying. 
Max-linear constructions, which replace the sum with the max operation  $A_{\max} \bm Z = \bigvee_{k = 1}^q \bm a_{\cdot k} Z_k$, are also multivariate regularly varying.
The angular measure of both models consists of $q$ point masses, whose locations are given by the $q$ normalized columns of $A$, which in turn result in the large points occurring on or near distinct rays as shown in Figure \ref{fig:max-lin_pointclouds}.
These linearly constructed models have been widely used because of their simple structure and interpretability.
In studying causal extremes, \cite{gissibl_kluppelberg_2018bernoulli} obtain the structure of a directed acyclic graph from the values in the matrix $A$.
In clustering for extremes, \cite{janssen_wan_22020kmeans} use these models and interpret the columns of $A$ as cluster centers. 
However, these linear and max-linear regularly varying models are not amenable to likelihood inference.
The distribution of the max-linear models contains a maximum operator and consequently the derivative (and therefore density) does not exist.
Perhaps more importantly, the scatterplots of data generated from these models are hard to reconcile with point clouds of actual data (e.g., Figure \ref{fig:late_ts_lag1}) whose large points do not lie along distinct rays.
In lieu of likelihood-based methods, practitioners have used alternative fitting methods, like least-squares estimators (see, e.g., \citealp{einmahl_kiriliouk_segers2018continuous}). 
% \DC{Is the denseness argument needed?}
% \citet{fougeres_mercadier_nolan2013dense} \strike{showed that any angular measure can be approximated arbitrarily well by the angular measure of a max-linear model if the number of point masses is large enough and thus any regularly varying model can be approximated by a max-linear model. }

%\DC{Now turn to TLETS}
The example which motivates this work is the family of transformed linear extremes time series (TLETS) models of \cite{mhatre_cooley2024}. 
TLETS models have ARMA-like structure and parameters, but are transformed-linear combinations of regularly varying noise (Section \ref{sec:tlets}). 
\citet{mhatre_cooley2024} assume regularly varying noise with tail index $\alpha = 2$ so that the tail pairwise dependence to behaves similarly to the autocovariance function (ACVF) of traditional ARMA models, and further use transformed-linear operations to guarantee a non-negative time series.
These features of the construction result in the model having an intractable likelihood as $\alpha = 2$ is the boundary for $\alpha$-stable models with heavy tails, and the transformed-linear arithmetic complicates analytic expressions.
Consequently, TLETS models have been fit with a moments-style method using an analogue to the innovations algorithm \citep{wixson_cooley2023attribution} and with least-squares-type methods (see, e.g., \citealp{mhatre_cooley2023_innov}).
Like the linear and max-linear models, although the TLETS models have familiar structure and interpretability, Figure \ref{fig:angular_measure_mismatch} shows that bivariate point clouds generated from the model may not resemble bivariate data.
For example, in Figure (\ref{fig:arma11_lag1}) mass is concentrated near the horizontal axis, near the identity ray, and near the ray with slope four.

%\DC{Now turn to what our method does.}
Our method involves obtaining an objective function from a second parametric family of models.
This objective function uses the likelihood from a proxy model and will be minimized like a (composite) likelihood. 
Like our target model, the proxy model will be MRV, and the target and proxy will share common univariate marginals.
The proxy model will be linked to our desired model through the TPD, which is a second-order summary of the bivariate dependence in any regularly varying model \citep{cooley_thibaud2019}. 
Importantly, although the objective function will be a product of pairwise likelihood expressions, this work differs from previous composite likelihood work in extremes in that we aim to find optimal values of the parameters from our target model rather than those of the proxy model.

% \DC{Second order focus}
By linking the models via the TPD, we focus on matching the second-order, bivariate dependence between the target and proxy models.
Focusing on second-order properties is a common practice in the statistics. 
Classical time series analysis assumes second-order stationarity to perform inference based on the ACVF.
Least-squares estimators in regression only require the (co)variance structure of the error terms to be specified.
Since neither time series nor regression require a model assumption, a Gaussian likelihood can be thought of as an objective function which provides information whether or not the errors are assumed to be Gaussian.
We view our method in the same framework. 
Recall that knowing the covariance in the Gaussian case means knowing the full dependence structure, but knowing covariance in the general setting does not fully describe the dependence in the joint distribution. 
This same idea holds when considering tail dependence as can be seen in Figure (\ref{fig:angular_measure_mismatch}). 
It will be seen below that the TPD determines the full dependence structure of our chosen proxy-likelihood. 

Our proxy model will come from the HR model because it has a continuous angular measure and an accessible closed form in the bivariate case.
The continuous angular measure of the HR model may match the data better than models with discrete angular measures. 
Figure (\ref{fig:hr_lag1_points}) is a scatter plot from a bivariate HR distribution that has the same TPD value as the fitted TLETS ARMA model at lag-$1$. 
Crucially, the HR distribution is parameterized by a dependence matrix with an entry $\lambda_{ij}^2$ for each pair of variables.
We will link the TPD values arising from our interpretable models to the $\lambda_{ij}$'s of the HR distribution.
While the HR distribution is fully characterized by these $\lambda_{ij}$'s, these parameters cannot be directly linked back to an interpretation of say cluster centers of a (max-)linear regularly varying model or the ARMA interpretation of a TLETS model.

The rest of this paper is as follows. 
In section \ref{sec:reg_var} we define regular variation and the TPD. 
In section \ref{sec:tlets} we review the desired models that we will use to demonstrate our method (TLETS models) and define their TPD functions. 
In section \ref{sec:proxy_lhood} we describe our proxy model, the TPD link, the composite likelihood approach, and derive the score function. 
 % and section \ref{sec:proxy_inference} discusses model selection. 
In section \ref{sec:censoring} we discuss two methods of ensuring large points inform about the tail dependence. 
Simulations are in section \ref{sec:proxy_sims} and an application to the wildfire data of \citep{wixson_cooley2023attribution} is in section \ref{sec:case_study}.

\section{Regular Variation}
\label{sec:reg_var}

We rely on the framework of MRV to model dependence in the upper tail. 
Regular variation is a common framework for extreme value theory (see, e.g., \citealp{dehaan_ferreira2006extreme, resnick2007heavy}) because both are naturally concerned with the tail of the distribution. 
One way to understand the tail of the distribution function $F$ of a random variable $X$ is to consider the rate of decay of the survival function $\bar{F}(x) = 1-F(x)$ as $x\rightarrow \infty$. 
The framework of regular variation naturally considers this decay. 

In the univariate case, regular variation has two common definitions. First, we say that $f$ is a regularly varying function {\em at infinity} if there exists an $\alpha \in \mathbb{R}$ such that $\lim_{t\rightarrow \infty} f(tx) / f(t) = x^{-\alpha}$. 
We denote this $f \in RV_\alpha$. 
Random variable $X$ is regularly varying if $\bar{F} \in RV_{\alpha}, \alpha > 0$. 
A second definition states that $\bar{F} \in RV_{\alpha}$ if and only if there exists a sequence $\{b_n\}$ with $b_n \rightarrow \infty$ such that 
$n\mathbb{P}( X / b_n \in \cdot ) \overset{v}{\rightarrow} \nu_\alpha(\cdot)$
where $\overset{v}{\rightarrow}$ denotes vague convergence in the space of nonnegative radon measures $M_+(0, \infty]$ and $\nu_\alpha(x,\infty] = x^{-\alpha}$ \citep[Theorem 3.6]{resnick2007heavy}. 

This second definition of univariate regular variation generalizes well to multiple dimensions.
\citet[Section 6]{resnick2007heavy} demonstrates that random vector ${\bf X}$ that takes values in $[0, \infty)^d$ is multivariate regularly varying if there exists a sequence $\{b_n\}$ with $b_n \rightarrow \infty$ and a Radon measure $\nu$ on the space $\mathbb{E} = [0, \infty]^d \setminus \{ {\bf0}\}$ such that in $M_+(\mathbb{E})$, $n\mathbb{P}( {\bf X} / b_n \in \cdot ) \overset{v}{\rightarrow} \bnu(\cdot).$ 
The limiting measure $\bnu$ has scaling property $\bnu(cB) = c^{-\alpha}\bnu(B)$ for any $c > 0$ and set $B \subset \mathbb{R}^d$. 
This scaling property leads to an independent polar decomposition of $\bnu$. 
Let $||\cdot||$ be a norm and define the unit ball $\mathbb{S}_{d-1} = \{ {\bf x} \in \mathbb{R}^d : ||{\bf x}|| = 1 \}$. 
Let $C(r, B) = \big\{ {\bf x} \in \mathbb{R}^d : ||{\bf x}|| > r, {\bf x} / ||{\bf x}|| \in B \big\}$ for some radial value $r > 0$ and Borel set $B \subset \mathbb{S}_{d-1}$. 
Then $\bnu [C(r, B)] = r^{-\alpha} H_{{\bf X}} (B)$ where $H_{{\bf X}} := \bnu [C(1, B)]$ is an angular measure for Borel sets in $\mathbb{S}_{d-1}$. 
This angular measure contains all of the dependence information. 
% Equivalently $\bnu ( dr \times d\bomega ) = \alpha r^{-\alpha-1}drdH_{{\bf X}}(\bomega)$. 
Finally, it should be noted that $b_n$, $\bnu$, and $H_{{\bf X}}$ are not uniquely determined as $b_n$ can be scaled by any positive constant and this will be absorbed into the limiting measure. 
Alternative to defining MRV on the positive orthant, MRV can be defined on random vectors $\mathbf{X}$ that take values in $\mathbb{R}^d$. We choose to define MRV such that we are looking at extremes in one direction. 

This definition of MRV holds for any finite dimension $d \geq 1$. 
The application of the method that we detail below is in a time series context. 
Time series are infinite dimensional random processes (i.e., $X_t$ takes values in $\mathbb{R}$ for $t = 1, 2, \dots$). 
We follow \citet{kulik_soulier2020} and say that a time series $\{X_t\}$ is regularly varying if all finite dimensional distributions are regularly varying with tail index $\alpha$ and the same scaling sequence $b_n$. 
Estimating or modeling the angular measure $H_{{\bf X}}$, which lies on the $d$-dimensional unit ball, becomes increasingly difficult as $d$ grows. 
It is for this reason that we consider summaries of the dependence. 

We follow \citet{cooley_thibaud2019} and many others (see, e.g., \citealp{mcgonigle2026moped} and references therein) and use the TPD as a summary measure of the pairwise dependence in any regularly varying random vector.
We define the TPD here in the time series context; it is simple to convert to the multivariate form. 
Let the time series $\{X_t\}$ be regularly varying with $\alpha = 2$ for all $t = 1, 2, \dots$ and tail stationary \citep{mhatre_cooley2024}; that is, the TPD function is a function of lag only. 
Define the two dimensional unit ball in the positive orthant $\mathbb{S}_{1}^+ = \{ {\bf x} \in \mathbb{R}^d : x_1, x_2 \geq 0 \text{ and } ||{\bf x}||_2 = 1 \}$ (where $|| \cdot ||_2$ is the Euclidean norm) and for each lag $h$, let the radial component $r_t = ||(x_t, x_{t+h})||_2$. 
The TPD function at lag $h$ is 
\begin{equation}\label{eq:tpd_def}
    \sigma(h) = \sigma(X_t, X_{t+h}) = \int_{ \mathbb{S}_1^+} s_1 s_2 dH_{X_t, X_{t+h}}({\bf s}), 
\end{equation}
where ${\bf s}$ is the angular component of $(X_t, X_{t+h})$: $s_1 = x_t / r_t$ and $s_2 = \frac{x_{t+h}}{r_t}$.

Although regular variation assumes the data are heavy tailed, it can be used as a dependence model for data which are not heavy tailed.
We will marginally transform data to regularly varying $\alpha = 2$ margins.
This idea is not uncommon in classical extreme value analysis where characterizations of the extreme value distributions are typically made assuming a particular marginal distribution \citep[][Section 6.1.2]{dehaan_ferreira2006extreme}, and is similar in spirit to transforming data to be plausibly Gaussian (such as applying a square root or logarithmic transformation) in order to fit classical statistical models.  
Our assumption that $\alpha = 2$ is made for convenience, \cite{kiriliouk2022} extended the definition of the TPD function for a general tail index, but its definition includes $\alpha$ in the integrand.

To estimate the dependence, we use the empirical TPD estimator of \citet{cooley_thibaud2019}. We reparameterize the lag-$h$ pairs of points $(x_t, x_{t+h}), \hspace{0.05in} t = 1, ..., n-h$ with polar coordinates. The radial component is the $L_2$-norm: $r_t = ||(x_t, x_{t+h})||_2$. The angular component ${\bf s} = (s_t, s_{t+h}) = (x_t, x_{t+h})/r_t$ places the point on the unit ball. The TPDF estimator under this parameterization is 

\begin{equation}
    \hat{\sigma}(h) 
    %= 2 \int_{\Theta_1^+} s_t s_{t+h} d\hat{N}_{X_t X_{t+h}}(s) 
    = \frac{2}{\sum_{t=1}^{n-h} \mathbb{I}(r_t > r_0)} \sum_{t=1}^{n-h} s_t s_{t+h} \mathbb{I}(r_t > r_0),
    \label{eq:tpdf_empirical}
\end{equation}
where the two arises from the known (after transformation) marginal distribution.
This estimator replaces the angular measure by its empirical estimate and considers only points above some high threshold $r_0$.

\section{Transformed Linear Extremes Time Series}
\label{sec:tlets}

The TLETS models of \citet{mhatre_cooley2024} are analogous to classical ARMA models with a few important distinctions (for an introduction to ARMA models see, e.g., \citealp{brockwell_davis2002}). 
Classical ARMA models are built out of linear combinations of white noise and it is common, but not necessary, to consider Gaussian noise. 
The TLETS models use transformed-linear combinations of regularly varying noise. 
Additionally, classical ARMA models are investigated based on the second-order property of covariance. 
TLETS models also consider a second-order property but in this case that property is the TPD. 

\subsection{Transformed-linear Operations}
\label{sec:trans_lin}

Extreme weather related wildfire risk events, like many other extreme events that we care about, occur in one direction in the tail. 
For this reason TLETS models were specifically designed to capture dependence in the upper tail (other regions like the joint lower tail can be modeled with simple transformations).
These models use transformed-linear operations \citep{cooley_thibaud2019} which are defined component-wise and involve a map, $f$, from the real line to the positive half line. 
For any two vectors in the positive orthant, ${\bf X}_1, {\bf X}_2 \in \mathbb{R}^d_+$, transformed-linear addition, denoted $\oplus$, is performed by mapping the components of ${\bf X}_1$ and ${\bf X}_2$ to the real line with $f^{-1}$, adding the two vectors, and then transforming the sum back to the half-line: ${\bf X}_1 \oplus {\bf X}_2 = f\{f^{-1}({\bf X}_1) + f^{-1}({\bf X}_2)\}$. 
Transformed-linear scalar multiplication, denoted $\circ$, works similarly: $a \circ \mathbf{X} = f\{a f^{-1}({\bf X})\}$. 
We use the function $f(x) = \log\{1 + \exp(x)\}$ as it has a negligible effect on the upper tail (i.e., $\lim_{x \rightarrow \infty} f(x) / x = 1$) and thus regular variation is preserved under these operations \citep{cooley_thibaud2019}.
Transformed-linear time series are constructed using transformed-linear operations in the place of classic arithmetic operations on a noise sequence $Z_t$ of independent, tail stationary, regularly varying $\alpha = 2$ random variables. 

In addition to using models constructed to ensure that the random process exists in the positive orthant, we restrict the definition of the TPD (\ref{eq:tpd_def}) to the positive orthant.
The TPD (\ref{eq:tpd_def}) is often defined as an integral on the entire unit ball $\mathbb{S}_{d-1}$ (see, e.g., \citealp{kiriliouk2022}). 
When defined in this way, the strength of the dependence in each region of the tail (e.g., jointly large positive or jointly large negative) is averaged with the dependence in the rest of the tail. 
This provides an additional complication when we are only interested in the tail in one direction and the dependence is not the same in, e.g., the upper and lower tails.

\subsection{Transformed-linear autoregressive model}

A time series $\{X_t\}$ is a transformed-linear auto regressive process of order 1 (denoted TL-AR(1)) if, for all t, 
\begin{equation}
    X_t = \phi \circ X_{t-1} \oplus Z_t, \quad Z_t \in RV_2^+
\end{equation}
where $|\phi| < 1$. 
When the marginal distribution of $X_t$ has scale 1 the TPD at lag-$h$ from a TL-AR(1) is 
\begin{equation}\label{eq:AR1_TPD}
    \sigma(h, \phi) = \max(0, \phi^h).
\end{equation}

\subsection{Transformed-linear moving average models}

We say that $\{X_t\}$ is a transformed-linear moving average process of order $q$ (denoted TL-MA($q$)) if, for all t, 
\begin{equation}
    X_t = \bigoplus_{j = 0}^q \theta_j \circ Z_{t-j}, \quad Z_t \in RV_2^+
\end{equation}
for $\theta_j \in \mathbb{R}$, $\theta_0 = 1$, and $\theta_q > 0$. The TPD at lag-$h$ from a TL-MA($q$) is $0$ whenever $h > q$. For $h \leq q$, When the marginal distribution of $X_t$ has scale 1, the TPD is 
\begin{align}\label{eq:MAq_TPD}
    \sigma(h, \theta_1, \dots, \theta_q) = & \frac{\sum_{l = 0}^q \theta_l^{(0)}\theta_{l+h}^{(0)}}{\sum_{l = 0}^q \theta_l^2}
\end{align}
where $a^{(0)} = \max(a, 0)$, and the last $h$ terms in the sum in the numerator are $0$. 

\subsection{Transformed-linear autoregressive moving average model}

The transformed-linear auto regressive moving average process of order $p=1$, $q=1$ (denoted TL-ARMA(1,1)) is given by 
\begin{equation}
    X_t \oplus  (-\phi) \circ X_{t-1}= Z_t \oplus \theta Z_{t-1}, \quad Z_t \in RV_2^+
\end{equation}
where $\phi, \theta \neq 0$ and $\theta + \phi \neq 0$. The TPD function for the ARMA(1,1) model is 
\begin{align}
    \sigma(h, \theta, \phi) = & \numberthis \label{eq:ARMA11_TPD}
    \begin{cases}
        \frac{\phi^{h-1}(\theta + \phi)(\phi \theta + 1)}{\theta^2 + 2 \phi \theta+ 1} & \text{if } \phi > 0, \phi + \theta > 0
        \notag\\
        0 & \text{if } \phi > 0, \phi + \theta < 0
        \notag\\
        \frac{\phi^h (\phi + \theta)^2}{(\phi + \theta)^2 - \phi^4 + 1} & \text{if } \phi < 0, \phi + \theta > 0, h \text{ is even}
        \notag\\
        \frac{\phi^{h-1} (\phi + \theta) (1 - \phi^4)}{(\phi + \theta)^2 - \phi^4 + 1} & \text{if } \phi < 0, \phi + \theta > 0, h \text{ is odd}
        \notag\\
        \frac{\phi^{h-1}(\phi + \theta)(\theta \phi^3 + 1)}{2\theta \phi^3 + \phi^2 \theta^2 + 1} & \text{if } \phi < 0, \phi + \theta < 0, h \text{ is even}
        \notag\\
        0 & \text{if } \phi < 0, \phi + \theta < 0, h \text{ is odd}.
    \end{cases} 
\end{align}

\subsection{Transformed-linear autoregressive moving average models of other orders}

\citet{mhatre_cooley2024} showed that a causal TL-ARMA model of any order can be represented by a TL-MA($\infty$) model. 
The TPD function of the TL-ARMA model is given by the coefficients of the TL-MA($\infty$) representation. 
The same is done in the computation of the auto-covariance function for a classical ARMA($p$, $q$) model in \citet{brockwell_davis2002}. 
In practice, if an ARMA(1,1) is not able to capture the dependence we can instead choose some large order MA model. 

\section{Defining our H\"usler-Reiss Composite Proxy-likelihood}
\label{sec:proxy_lhood}

\subsection{The H\"usler-Reiss distribution}\label{sec:hr}
It is well known that properly normalized block maxima of bivariate normal random variables with any constant covariance less than one converge in distribution to a separable bivariate max-stable model (see e.g., \citealp{Geffroy_1958, Sibuya_1960}).
\citet{husler_reiss1989maxima} constructed a non-degenerate max-stable distribution from block maxima of normal random variables by making the covariance $\rho$ increase with the sample size. 
If the covariance $\rho(n)$ satisfies $[1-\rho(n)]\log(n) \rightarrow \lambda^2 \in [0, \infty]$ as $n \rightarrow \infty$ then the distribution of block maxima is given by 

\begin{equation}\label{eq:HR_df_Gumbel}
    F_{\lambda}(x_1, x_2) = \exp \left\{ - e^{-x_1} \Phi \left( \lambda + \frac{x_2 - x_1}{2\lambda}\right) - e^{-x_2} \Phi \left( \lambda + \frac{x_1 - x_2}{2\lambda} \right) \right\},
\end{equation}
where $\Phi(\cdot)$ is the standard normal distribution function. 

\citet{husler_reiss1989maxima} extend the result to the multivariate case. 
As with all multivariate max-stable models, the distribution function can be written as $\exp[-V(\bf{x})]$ where $V(\bf{x})$ is the exponent measure.
In this case $V(\bf{x})$ is in the form of a sum of a function of lower-dimensional margins. 
The sum is indexed by all possible combinations of indices from lower-dimensional margins.
Let $\Lambda = \{\lambda_{ij}^2\}_{1 \leq i, j \leq d}$ be the the symmetric, conditionally negative definite matrix which parameterizes a $d$-dimensional HR distribution. 
Let ${\bf m} = m_0, \dots, m_l$ where $l \in 1, \dots, (d-1)$ and $0 \leq m_0 < \dots <m_l \leq d-1$ such that $\bf{m}$ indicates which margins to include and $l$ indicates the number of included indices (i.e., the lower dimension). The distribution function of the $d$-dimensional HR distribution, on Gumbel margins, is given by 

\begin{align}\label{eq:multi_HR}
    F_{\Lambda}({\bf x}) = \exp\Bigg\{ \sum_{l = 0}^{d-1} \hspace{0.1in} \sum_{{\bf m} : 0 \leq m_0 < \dots <m_l \leq d-1} f_{l,{\bf m}, \Lambda} (x_{m_1}, \dots, x_{m_l}) \Bigg\}.
\end{align}
Here $f_{l,{\bf m}, \Lambda}$ is the $l$-dimensional component of the exponent measure that includes margins ${\bf m}$. 
For the full form of the multivariate HR distribution see \citet{husler_reiss1989maxima} or \citet{engelke_malinowski_kabluchko_schlather_2015estimation}. 
Equation (\ref{eq:multi_HR}) makes it clear that, as the dimension increases, the number of terms in the exponent measure grows combinatorially. 
This result, and the multivariate integrals required, demonstrates that the likelihood is onerous even at reasonably small dimensions. 

\citet{engelke_malinowski_kabluchko_schlather_2015estimation} showed that, when the data are in the maximum domain of attraction of $F_\Lambda$, the distribution of so-called extremal increments is multivariate normal. 
They use this normal likelihood to perform inference. 
Their definition of extremal increments requires conditioning on a fixed component being large which does not naturally fit our use case. 

We rely on the closure property of the HR distribution; the lower dimensional margins of the HR distribution are HR distributions with dependence parameters which are equal the respective components of $\Lambda$. 
This allows us to follow \cite{padoan_ribatet_sisson2010} and use a bivariate composite likelihood approach. 
As such, we only consider the bivariate HR distribution.

The TLETS models are defined on regularly varying $\alpha = 2$ margins so we transform the margins of the HR distribution (\ref{eq:HR_df_Gumbel}) to be Frechet($2$). 
Let $G_{\lambda_{ij}}(x_i, x_j) = F_{\lambda_{ij}}(\log(x_i^2), \log(x_j^2))$ which has Frechet(2) margins and is given by
% Frechet(2) HR distribution
\begin{align}
    G_{\lambda_{ij}}(x_i, x_j) =  \exp \left[ - x_i^{-2} \Phi \left\{   \lambda_{ij} - \frac{ 2\log \left(\frac{x_i}{x_j} \right)}{\lambda_{ij}} \right\} - x_j^{-2} \Phi \left\{ \lambda_{ij} - \frac{2\log \left(\frac{x_j}{x_i} \right)}{\lambda_{ij}} \right\} \right] 
    =   \exp \left\{ -V(x_i, x_j, \lambda_{ij}) \right\}.
\end{align}
Leaving out the arguments for the exponent measure $V$, the density is 
% density of Frechet(2) HR dist
\begin{align}\label{eq:HR_density_lambdas}
    g_{\lambda_{ij}}(x_i, x_j) = \frac{\partial^2}{\partial x_i \partial x_j} G_{\lambda_{ij}}(x_i, x_j) = G_{\lambda_{ij}}(x_i, x_j) \bigg( \frac{\partial}{\partial x_i} V \frac{\partial}{\partial x_j} V - \frac{\partial^2}{\partial x_i \partial x_j} V \bigg)
\end{align}
where the derivatives in (\ref{eq:HR_density_lambdas}) are in Appendix~\ref{ap:hr_model}.

\subsection{The tail pairwise dependence link}
\label{sec:hr_tpd_link}

The basis of our method is the link between the $ij^{th}$ parameter of the HR distribution ($\lambda_{ij}$) and the $ij^{th}$ TPD parameter ($\sigma_{ij}$) which is defined in (\ref{eq:tpd_def}). In appendix~\ref{ap:tpd_link} we obtain the angular measure $h_{\lambda_{ij}}$ from the exponent measure $V$. Let $s_i = x_i / ||\mathbf{x}_{ij}||_2$ be the first dimension of the pseudo-angular component of the vector $\mathbf{x}_{ij}$, the angular measure is 
\begin{align}
    h_{\lambda_{ij}}(s_i) = & \frac{1}{2 \sqrt{1-s_i^2}}\Bigg\{ \frac{\lambda_{ij}^2 + \log\frac{s_i}{\sqrt{1-s_i^2}}}{\lambda_{ij}^3 s_i^3 \sqrt{1-s_i^2}} \phi \left(\lambda_{ij} - \frac{ \log \frac{s_i}{\sqrt{1-s_i^2}}}{\lambda_{ij}}\right) + \frac{\lambda_{ij}^2 + \log\frac{\sqrt{1-s_i^2}}{s_i}}{\lambda_{ij}^3 s_i \sqrt{1-s_i^2}^3 } \phi \left(\lambda_{ij} - \frac{\log \frac{\sqrt{1-s_i^2}}{s_i}}{\lambda_{ij}} \right) \Bigg\}. 
\end{align}
The $ij^{th}$ parameter of the HR distribution has $ij^{th}$ TPD parameter
\begin{align}\label{eq:HR_TPD}
    \sigma_{ij} = & \int_0^1 s_i \sqrt{1-s_i^2} h_{\lambda_{ij}}(s_i) ds_i = \exp\left(-\frac{\lambda_{ij}^2}{2}\right).
    %\notag\\ 
    %= & \frac{1}{2} \int_0^1 s_i \Bigg\{ \frac{\lambda_{ij}^2 + \log\frac{s_i}{\sqrt{1-s_i^2}}}{\lambda_{ij}^3 s_i^3 \sqrt{1-s_i^2}} \phi \left(\lambda_{ij} - \frac{ \log \frac{s_i}{\sqrt{1-s_i^2}}}{\lambda_{ij}}\right) +
    %\frac{\lambda_{ij}^2 + \log\frac{\sqrt{1-s_i^2}}{s_i}}{\lambda_{ij}^3 s_i \sqrt{1-s_i^2}^3 } \phi \left(\lambda_{ij} - \frac{\log \frac{\sqrt{1-s_i^2}}{s_i}}{\lambda_{ij}} \right) \Bigg\} ds_i.
\end{align}

\begin{figure}
    \centering
    \includegraphics[width=0.6\linewidth]{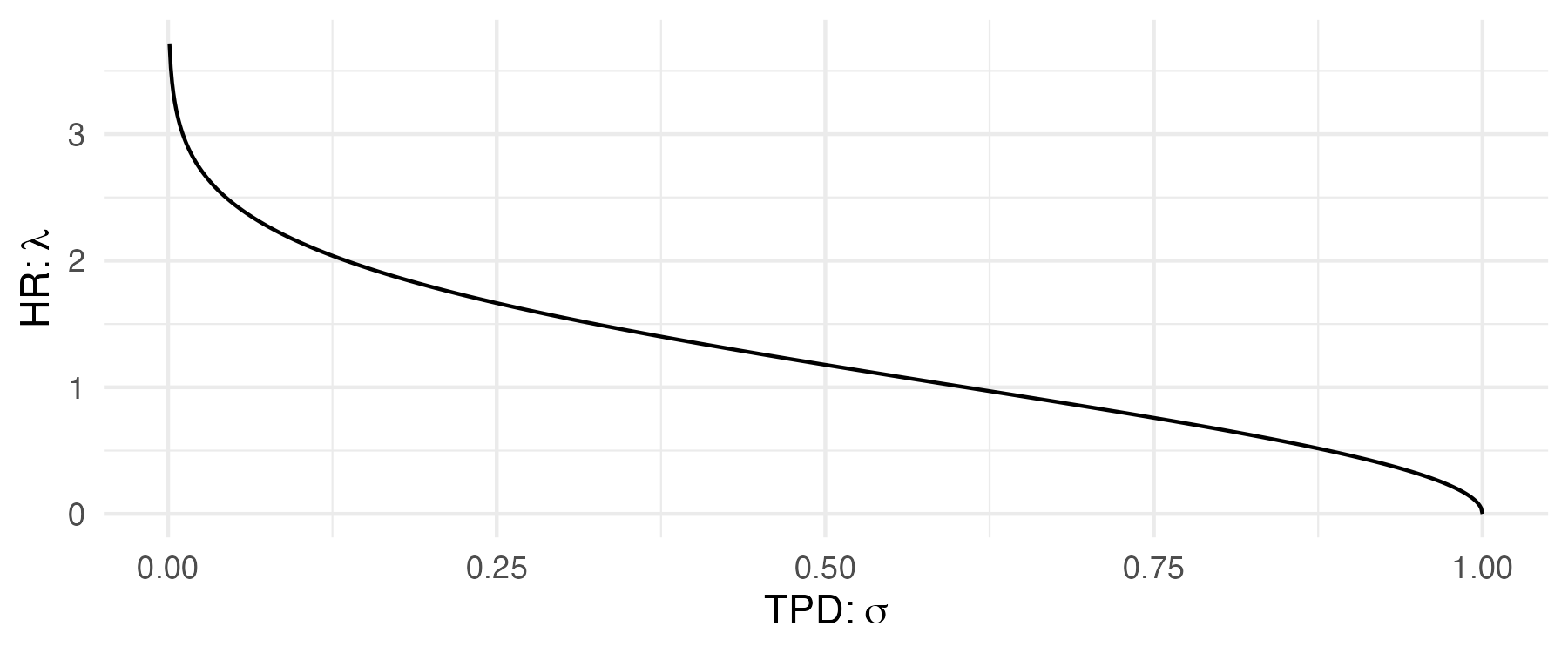}
    \caption{Link between HR dependence parameter ($\lambda$) and TPD dependence parameter ($\sigma$).}
    \label{fig:HR_TPD_func}
\end{figure}

\begin{figure}
    \centering
    \subfloat[Weak Dependence]{\label{fig:HR_low_dep}\includegraphics[width=0.35\textwidth]{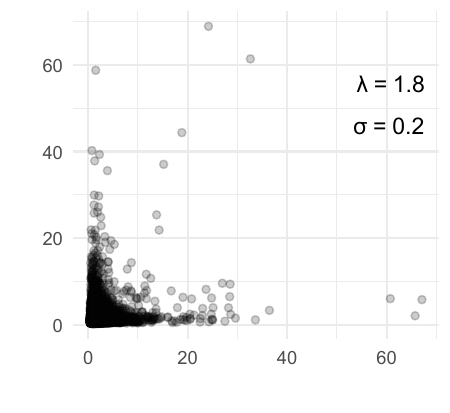}}
    \subfloat[Strong Dependence]{\label{fig:HR_high_dep}\includegraphics[width=0.35\textwidth]{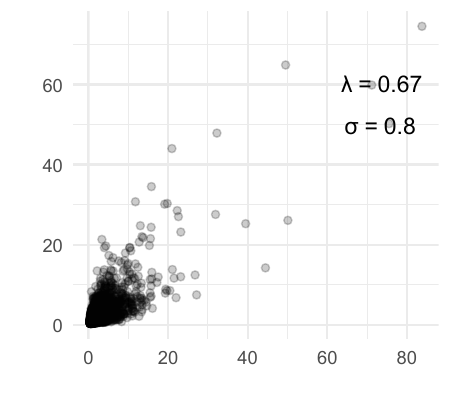}}
    \caption{HR point clouds (10000 points) with dependence parameter ($\lambda$) and TPD parameter ($\sigma$).}
    \label{fig:HR_pointclouds}
\end{figure}

The integral in (\ref{eq:HR_TPD}) is computed in Appendix~\ref{ap:hr_tpd}. 
Let $\lambda(\cdot)$ be the inverse function which takes a TPD value and returns the bivariate HR dependence parameter (called $\lambda_{ij}$ up until this point) and let $\Lambda(\cdot)$ be the matrix version of $\lambda(\cdot)^2$. 
Figure \ref{fig:HR_TPD_func} plots the function $\lambda(\sigma)$ which demonstrates the bijection between the TPD and the dependence parameter of a bivariate HR distribution. 
Figure \ref{fig:HR_pointclouds} shows an example HR point cloud under weak and strong dependence respectively. 
Each point cloud has 10000 points. 
Combining the TPD function from one of the TLETS models (\ref{eq:AR1_TPD}), (\ref{eq:MAq_TPD}), or (\ref{eq:ARMA11_TPD}) with $\lambda(\cdot)$ gives a map between the TLETS parameters and the HR parameters that have the equivalent TPD. 
Let $\btheta$ be the parameter vector for the TLETS model that is being fit (e.g., a TL-ARMA(1,1) model has $\btheta = (\phi, \theta)$).
This map is the link between the two models and results in a bivariate HR distribution which can be written explicitly as a function of the TLETS parameters: 

\begin{align}\label{eq:HR_density}
    g_{ \lambda\{\sigma(h, \btheta)\} }(x_n, x_{n+h}) = & G_{ \lambda\{\sigma(h, \btheta)\} } (x_n, x_{n+h}) \bigg( \frac{\partial}{\partial x_n} V \frac{\partial}{\partial x_{n+h}} V - \frac{\partial^2}{\partial x_n \partial x_{n+h} } V \bigg) 
    \\
    \text{where } V = & V\big\{x_n, x_{n+h}, \lambda[\sigma(h, \btheta)]\big\}.
    \notag
\end{align}

\subsection{The H\"usler-Reiss composite proxy-likelihood}
\label{sec:composite_lhood}
Our method combines the pieces shown above using a composite likelihood approach. 
Composite likelihood approaches go as far back as \citet{lindsay1988composite} and are a useful alternative to full likelihood inference when the full likelihood is analytically unavailable or computationally infeasible (see, e.g., \citealp{varin2011overview}). The idea is to use a combination of valid likelihoods (e.g., bivariate or conditional likelihoods) in the place of the full likelihood. We follow \citet{padoan_ribatet_sisson2010} and use the bivariate margins as our likelihood components. 

We adapt the composite likelihood to the time series context in the following manner. 
Let $\btheta$ be the set of parameters for the full target model likelihood. Let ${\bf x_i} = (x_{i+1}, x_{i+2}, \dots, x_{N})^T$ where $N$ is the length of the observed time series. 
Let $L_h$ be the likelihood for the lag-$h$ margin and $w_h$ be user specified weights. The bivariate composite likelihood for time series is

\begin{equation}
    L_{cl}(\btheta|\mathbf{x}) = 
    % \prod_{h = 1}^{\infty}  L_{h}^{w_{h}}(\btheta|\mathbf{x}^h) = 
    \prod_{h = 1}^{N-1} \prod_{n = 1}^{N-h} [L_{h}(\btheta|x_n, x_{n+h})]^{w_{h}}.
\end{equation}
Here the first product is over lags (i.e., the bivariate margins we are considering) and the second product is over the $N-h$ observations of lag-$h$ points. 
Time series are infinite dimensional processes and thus the number of terms in the first product is limited by the number of observations.
In this work we consider some large maximum lag value $h_{max}$ which is equivalent to so-called tapered weights which set $w_h := 0, \forall h > h_{max}$. 
A user could consider other weights such as weights which are inversely proportional to the lag (e.g., $h^{-1}$) but we have not found that useful in simulations and thus we consider unit weights for all $h < h_{max}$ in the following. 
When using the method for model selection, $h_{max}$ must be the same in all model fits.

A composite likelihood approach is necessary as the analytic form of the $h_{max}$-dimensional HR likelihood becomes unwieldy for even moderate dimensions.
Additionally, constructing the objective function via composite bivariate HR likelihoods allows additional flexibility which may not be immediately apparent.
In the finite dimensional setting, any valid (i.e., completely positive) TPD matrix corresponds to valid MRV models \citep{cooley_thibaud2019}.
The Gaussian-based construction of the multivariate HR model imposes the additional restriction that the matrix of $\lambda_{ij}^2$'s must be conditionally negative definite.
It is possible to have a valid collection of TPD $\sigma_{ij}$'s which correspond to an invalid collection of HR $\lambda_{ij}^2$'s.
Specifically, in terms of TLETS models, we are able to find valid TL-ARMA(1,1) parameter combinations $(\phi, \theta)$ which, after passing through the TPD-to-HR relationship given by the inverse of (\ref{eq:HR_TPD}), result in $\lambda_{ij}^2$'s that do not assemble into a conditionally negative definite matrix.
By constructing an objective function which is composed from bivariate HR likelihoods, our method does not constrain the TPD (or TLETS) parameter space to the restricted subspace corresponding to valid multivariate HR models.

Replacing $L_h$ with the bivariate HR density (\ref{eq:HR_density}) the composite log-likelihood is
% Composite log-likelihood
\begin{align}\label{eq:biv_cllhood}
    \ell_{cl} (\btheta | \mathbf{x}) 
    = & \sum_{h = 1}^{h_{max}} \sum_{n = 1}^{N-h} w_{h} \log g_{ \lambda\{\sigma(h, \btheta)\} }(x_n, x_{n+h})
    \notag\\
    = & \sum_{h = 1}^{h_{max}} w_h \sum_{n = 1}^{N-h} \Bigg\{ -V + \log \bigg( \frac{\partial}{\partial x_n} V \frac{\partial}{\partial x_{n+h}} V  - \frac{\partial^2}{\partial x_n \partial x_{n+h}} V \bigg) \Bigg\}
\end{align}
where the derivatives are defined in Appendix~\ref{ap:hr_model}. Equation (\ref{eq:biv_cllhood}) is what we term the bivariate HR composite proxy-likelihood (proxy-likelihood for short).

\subsection{The score function} 

The contribution of one point (i.e., one $(x_n, x_{n+h})$ pair) to the score function for some $\theta_k \in \btheta$ is, by the chain rule, of the form 
\begin{align}\label{eq:score_contrib}
    \frac{\partial}{\partial \theta_k} \ell_{cl}[\lambda\{\sigma(\btheta)\} | x_n, x_{n+h}] = & \frac{\partial}{\partial \lambda_h} \ell_{cl}(\lambda_h | x_n, x_{n+h}) \frac{\partial}{\partial \sigma} \lambda(\sigma) \frac{\partial}{\partial \theta_k} \sigma(h, \btheta).
\end{align}
Obtaining the first and third factors is done in Appendix~\ref{ap:score_func}. The middle factor is $\frac{\partial}{\partial \sigma} \lambda(\sigma) = -\left[\sigma \sqrt{-2 * \log(\sigma)}\right]^{-1}$. 

Notice that, for $h \neq h'$, lag-$h$ and lag-$h'$ points will both contribute to the score for $\theta_k$ if the third factor, $\frac{\partial}{\partial \theta_k} \sigma(\cdot, \btheta)$, is non-zero for $h$ and $h'$. Finally, note that the third factor in (\ref{eq:score_contrib}) is model dependent as this is the map between the TLETS parameters and the TPD. Application of our proxy-likelihood method to different models in multivariate extremes requires computing this component of the score function. 

The first factor in (\ref{eq:score_contrib}) is the only component that includes the data.
It is in this component that the data informs the HR likelihood surface. 
We demonstrate in Section~\ref{sec:tpd_estimation} that we can ignore the other components of the score and use the maxima of the HR likelihood surface as a TPD estimator due to the one-to-one mapping (\ref{eq:HR_TPD}). 
Simulations in Section~\ref{sec:tpd_estimation} suggest that this estimator is less biased (especially in later lags) than the empirical estimator (\ref{eq:tpdf_empirical}). 
It would be possible to use these likelihood-based estimated TPD values in lieu of estimates from (\ref{eq:tpdf_empirical}) the for fitting TLETS models using the moments-based or least squares methods in prior work \cite{wixson_cooley2023attribution, mhatre_cooley2024}.
Instead, we will prefer to use the weighted sum of all three components of the score function so that we optimize our objective function over the space of valid TLETS parameters; thereby directly integrating the TLETS parameters into the optimization. 
Constraining the space with the third factor in the score function ensures that our resulting TPD function is a positive semi-definite function as all valid TLETS parameters define valid TPD functions.

\section{Censoring Approach to Parameter Estimation}
\label{sec:censoring}

Key to any analysis of extreme events is the idea that extreme data are needed to inform about the tails of the process. We seek to retain as much information as possible while at the same time ignoring the information in the bulk of the distribution that would cause bias. Two common approaches available to bivariate regular variation are Euclidean censoring (see, e.g., \citealp{smith_tawn_coles1997markov} and \citealp{huser_davison_genton_2016}) and only including data whose radial component exceeds a high threshold. This second approach is commonly done in the estimation of the TPD (see, e.g., \citealp{cooley_thibaud2019}). We briefly explore these two censoring schema as our method can be employed with either. 

\citet{huser_davison_genton_2016} show that using Euclidean threshold censoring in pairwise likelihood estimation has lower bias and RMSE than other proposed likelihood estimators. The central idea behind this censoring is to transform the joint distribution of the data into a joint distribution of threshold exceedance indicators and threshold exceedances. This joint distribution treats values smaller than the threshold as censored. By essentially counting the small points (rather than discarding them) this method retains some information from every observation point. We follow \cite{smith_tawn_coles1997markov} and develop this joint distribution in Appendix~\ref{ap:censoring}.

Radial censoring is employed in, e.g.,  \citet{jiang_cooley_wehner_2020principal, wixson_cooley2023attribution}. 
Relying on the asymptotic independence between the radial and angular components in regularly varying random vectors, this method ignores all points which are not large when estimating the tail dependence. 
Consider $(X_i, X_j)\in RV_{\alpha}$ and transform to pseudo-polar coordinates as was done in the derivation of the TPD from an HR distribution: let $R = ||(X_i, X_j)||_2 = \sqrt{X_i^2 + X_j^2}$ and $\mathbf{S} = (S_i, S_j) = (X_i, X_j)/R$ so that $R$ is the radial component of each point and $\mathbf{S}$ is the pseudo-angular component. 
Radial censoring involves fixing some large quantile $r_0$ of the radial distribution and including points in the analysis only if the radial component is larger than $r_0$. 

\begin{figure}
    \centering
    \subfloat[Weak Dependence]{\label{fig:radEuc_weak}\includegraphics[width=0.8\textwidth]{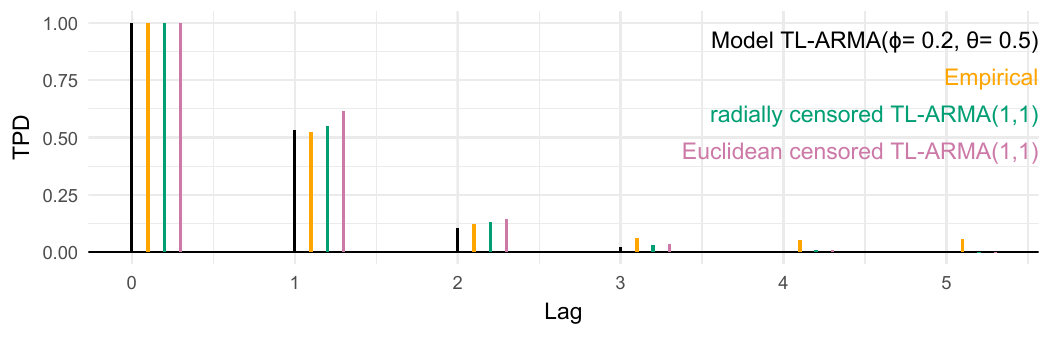}}
    \\
    \subfloat[Strong Dependence]{\label{fig:radEuc_strong}\includegraphics[width=0.8\textwidth]{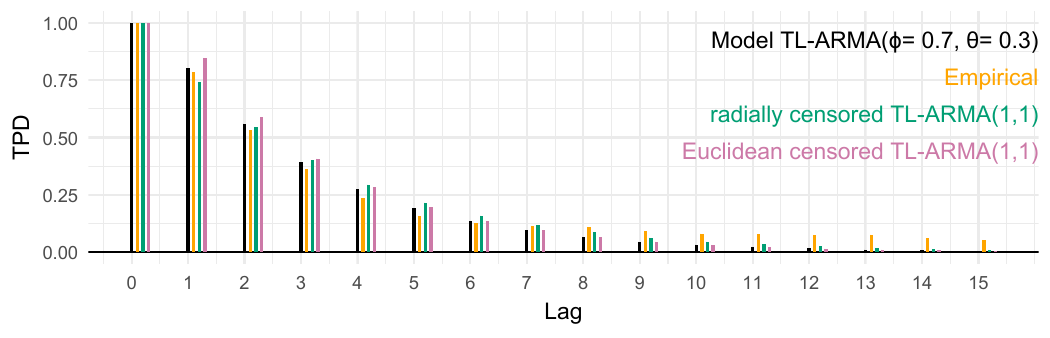}}
    \caption{Comparison of radial and Euclidean censoring techniques. The model, empirically estimated, radially censored and Empirically censored TPDs are plotted in black, yellow, green, and pink respectively.}
    \label{fig:radEuc_comparison}
\end{figure}

While both approaches have their strengths we found that the simpler radial approach which closely matches the estimation of second order dependence (our target) often performs better and is less computationally expensive. 
Figure (\ref{fig:radEuc_comparison}) shows two examples of time series of length 10000 which were fit using the two different censoring schemes.
In general, the radially censored model fits the early lags better (the smaller fitted TPD value is closer to the true TPD value) and the Euclidean censored model decays to zero faster. 

\section{Simulations}
\label{sec:proxy_sims}

We demonstrate our proxy-likelihood in the following simulations. In each simulation below we simulate time series of length 10000 from the model specified, use the 0.95 radial quantile as the censoring threshold, set $h_{max} = 20$, and use 20 sub-series for estimating the covariance of the score when needed for information-criteria based model selection (see Appendix~\ref{ap:model_selection}). Each simulation is repeated 100 times to assess the variability.

\subsection{Proxy-likelihood as a tail pairwise dependence estimator}\label{sec:tpd_estimation}
\begin{figure}
    \centering
    \includegraphics[width=0.95\linewidth]{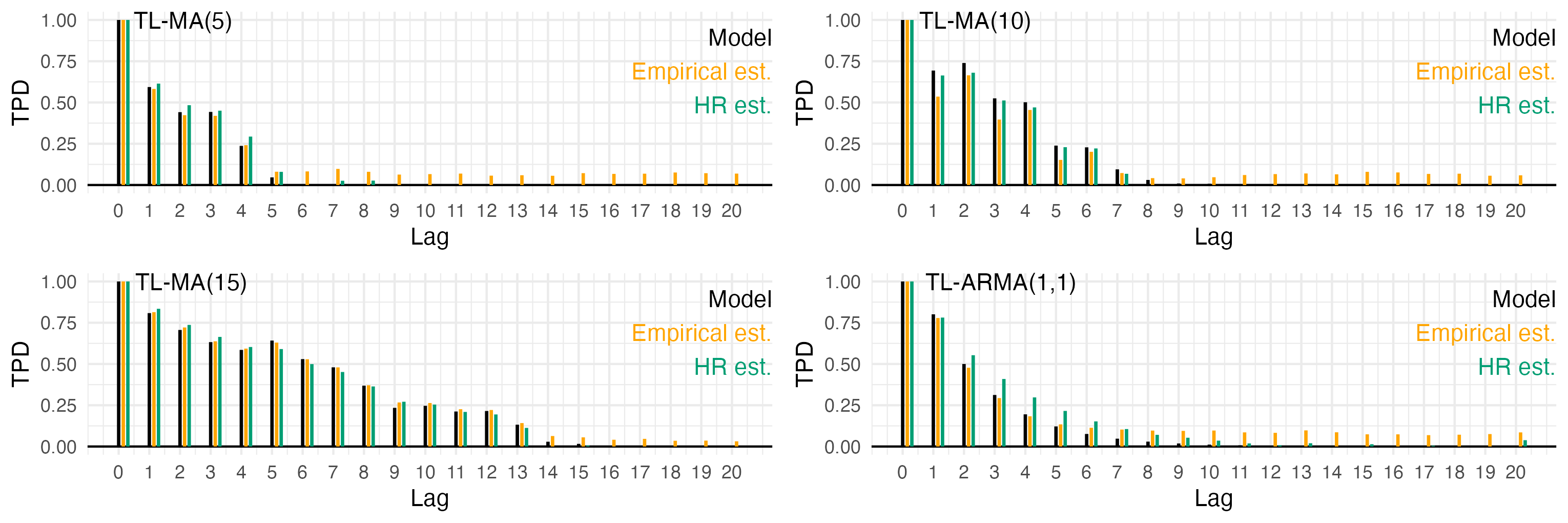}
    \caption{Model TPD (black) compared to the TPD estimated with the empirical estimator (yellow) and the TPD estimated with the HR likelihood (green). The plot in the top-left is from a TL-MA(5), in the top-right is a TL-MA(10), in the bottom-left is a TL-MA(15), and in the bottom-right is a TL-ARMA(1,1).}
    \label{fig:tpd_est_comparison}
\end{figure}

Our proxy-likelihood method is centered on the link between the HR TPD and the TPD of the TLETS models. 
We verify the mapping between the HR parameters and the TPD with simulated data from a variety of TLETS models. 
We use the HR composite likelihood as a TPD estimator by removing the last two factors from the score function (\ref{eq:score_contrib}). 
We thus compute the TPD values from the fitted HR parameters ($\hat{\lambda}_{ij}$'s) from simulated time series. 
We compare the HR-estimated TPD values to the model TPD values to see if the proxy model captures the second-order dependence. 
The empirical TPD estimator (\ref{eq:tpdf_empirical}) is used as a baseline for comparison as it is the existing TPD estimator. 
%We ensure that the estimated matrix of TPD values is positive definite with the \texttt{nearPD()} R function in both estimation methods. 
Four of these simulations (from a TL-MA(5), TL-MA(10), TL-MA(15), and TL-ARMA(1,1)) are shown in Figure~\ref{fig:tpd_est_comparison}. 
These simulations demonstrate that the HR TPD estimator is able to capture the TPD.

In all cases the HR TPD estimator does a better job of recognizing when the model dependence has decayed to zero whereas the empirical estimator has more bias in these later lags. 
This bias arises from estimating an asymptotic quantity with a finite number of points; the radial quantile used as a censoring threshold should be allowed to go to 1 but it must be fixed at some level strictly less than 1 if we are to have any points to use in the estimation of the TPD. 
In this way we have a classic bias-variance tradeoff in the estimation of the TPD. 
The empirical estimator inherently assumes we are in the asymptotic tail where the angular measure exists and thus we discard all radial information in the estimation of the TPD. 
By contrast, the HR proxy-likelihood retains the radial information and uses that information in estimating the dependence parameter $\lambda$ (and therefore the TPD). 
The HR copula is determined by $\lambda$ at all radial distances and thus our proxy-likelihood is able to learn from more information than the empirical estimator. 
We caution that our estimator still suffers from the bias introduced through using a tail model for points that are not in the asymptotic tail.  

We assess the HR TPD estimator in a more formal fashion by computing the sum (across lags) of absolute difference from the model TPD on 100 simulated time series from each model. 
We choose to compare methods using the sum of absolute differences (as opposed to squared differences) as the magnitude of errors is more directly interpretable.
We compare the estimation error in the HR TPD estimator to that of the empirical estimator (\ref{eq:tpdf_empirical}). 
First, we include all lags up to some fixed large value (here we use 20) in the sum as this more closely resembles the case where we do not know the generating model and balances capturing true dependence with limiting the bias at later lags.
Second, we only include lags with non-zero model TPD in the sum to assess whether the estimator can capture the dependence present in the model. 
TL-ARMA models have non-zero dependence at all lags but that dependence decays exponentially with lag and thus the dependence becomes negligible at some lag. 
In the second case, we only include lags where the model dependence is greater than 0.01 for the TL-ARMA models. 
The results from this series of simulations (Table~\ref{tab:tpd_est_results}) agree with our observations from Figure~\ref{fig:tpd_est_comparison}.
In particular, the HR TPD estimator has lower mean (across simulations) error when including all lags and in these cases the HR estimator has smaller error in 97, 100, 84, and 83 out of 100 simulations (depending on the model). 
The empirical estimator has lower mean error when only including lags with non-negligible TPD and has smaller error in 77, 52, 43, and 77 out of 100 simulations (depending on the model). 

\begin{table}
\centering
\begin{tabular}{lrrr}
 &  \multicolumn{3}{c}{Lags 1-20 included in sum} \\ 
  \hline
Model & mean (sd) HR & mean (sd) empirical & prop HR better \\
   \hline
  TL-MA(5) & 0.36 (0.33) & 1.17 (0.15) & 0.97  \\ 
  TL-MA(10)  & 0.46 (0.26) & 1.03 (0.23) & 1.00  \\ 
  TL-MA(15) & 0.58 (0.29) & 0.89 (0.24) & 0.84   \\
  TL-ARMA(1,1) & 0.47 (0.41) & 1.09 (0.29) & 0.83 \\ 
   \hline
    & \multicolumn{3}{c}{Lags with non-negligible TPD included in sum} \\ 
   \hline
  TL-MA(5) &   0.19 (0.09) &  0.12 (0.05) & 0.23 \\ 
  TL-MA(10) &   0.34 (0.15) &  0.33 (0.12) & 0.48  \\ 
  TL-MA(15) &   0.50 (0.22) &  0.55 (0.21) & 0.57   \\
  TL-ARMA(1,1) &   0.38 (0.37) &  0.27 (0.28) & 0.23    \\ 
\end{tabular}
\caption{Mean (sd) of sum of absolute error in TPD estimation from 100 time series simulated from each of the four listed models.} 
\label{tab:tpd_est_results}
\end{table}

\subsection{Proxy-likelihood model fitting}\label{sec:proxy_model_fitting}

While Section 6.1 focuses on whether the proxy-likelihood approach can be used to estimate TPD values, here we assess whether the proxy likelihood, when used to estimate TLETS model parameters, accurately captures the true lagged dependence.
We use all simulated time series from the previous simulation (Section~\ref{sec:tpd_estimation}) and fit the generating model with our proxy-likelihood method. 
The sum of absolute error from the model TPD is used to compare our proxy-likelihood fitted model to a model fitted by an existing method. 
As existing methods require an estimate of the TPD; we use the empirical estimator (\ref{eq:tpdf_empirical}) as that is what existed in the literature prior to this work. 

For the MA models, we compare the proxy-likelihood fit to the fit resulting from the innovations algorithm as was done in \cite{wixson_cooley2023attribution}.
Results are shown in Table~\ref{tab:model_fit_results}.
When fitting the generating model, fitting by the proxy-likelihood reduces the error when compared to the innovations algorithm by 29\%, 44\%, and 47\% for the MA(5), MA(10), and MA(15) models respectively.
We additionally fit MA(20) models to all three generated models, as due to the bias issues of the empirical TPD estimator, it can be difficult to determine the order of an MA model from TPD values estimated by (\ref{eq:tpdf_empirical}).
Table~\ref{tab:model_fit_results} shows that when fitting this misspecified model, the error of the TPD estimates is also reduced by using the proxy likelihood approach over the innovations algorithm.
The proportion of simulations where the proxy likelihood estimation method outperformed the innovations algorithm was roughly 80\%, except for when fitting the generating model of the MA(5) where this proportion was 66\%.

For the ARMA(1,1), we compare the proxy likelihood approach to an approach which minimizes the difference between the empirically-estimated lagged TPD, and the lagged TPD of a fitted ARMA(1,1) model.
We perform two comparison fits; the first reflects what was done in previous literature \citep{mhatre_cooley2024} by finding which parameters minimize the sum of squared errors to the empirical TPD, the second recognizes that the metric for comparison in the current study is least-absolute error and thus we find which parameters minimize the sum of absolute errors to the empirical TPD.
Here the results are a little less conclusive, as shown in Table~\ref{tab:model_fit_arma_results}.
The proxy likelihood method has a mean absolute error of 0.40, which is less than the either of the comparison methods which have errors of 0.48 and 0.42.  
However, the proportion of simulations in which the proxy likelihood method was found to have less error is less than half.

\begin{table}
\centering

\begin{tabular}{lrrr}
 &  \multicolumn{3}{c}{When fitting the generating model} \\ 
  \hline
Model & mean (sd) proxy & mean (sd) innovations & prop HR better \\
   \hline
  TL-MA(5) &   0.20 (0.09) &  0.28 (0.11) & 0.66   \\ 
  TL-MA(10) &   0.32 (0.15) &  0.57 (0.20) & 0.80   \\ 
  TL-MA(15) &   0.48 (0.24) &  0.91 (0.39) & 0.82    \\
   \hline
    & \multicolumn{3}{c}{When fitting a TL-MA(20)} \\ 
   \hline
  TL-MA(5) & 0.52 (0.41) & 1.02 (0.14) & 0.80  \\ 
  TL-MA(10)  & 0.49 (0.32) & 0.79 (0.19) & 0.83 \\ 
  TL-MA(15) & 0.57 (0.34) & 0.79 (0.32) & 0.74  \\
\end{tabular}
\caption{Mean (sd) of sum of absolute error in TPD from model fitting to 100 time series simulated from each of the three listed models. We use ``proxy'' to indicate our method, ``innovations'' to indicate the existing method of fitting using the extremes analogue to the innovations algorithm.} 
\label{tab:model_fit_results}

\begin{tabular}{lrrrrr}
 & mean (sd) proxy & mean (sd) LS & prop HR better LS & mean (sd) LA  & prop HR better LA \\ 
  \hline
 TL-ARMA(1,1) &  0.40 (0.40) &  0.49 (0.57) & 0.48 & 0.42 (0.57)  & 0.39 \\
   \end{tabular}
   \caption{Mean (sd) of sum of absolute error in TPD from model fitting to 100 time series simulated from TL-ARMA(1,1). We use ``proxy'' to indicate our method, ``LS'' to indicate the existing method of fitting least-squared error from the estimated TPD, and "LA" to indicate fitting with least-absolute error.} 
   \label{tab:model_fit_arma_results}
\end{table}

\subsection{Proxy-likelihood model selection}\label{sec:proxy_model_selection}
Our method is built on a second-order summary of the joint distribution and thus targets parameters which best capture this summary of the tail dependence. 
The previous two simulation studies demonstrate that our method achieves these goals. 
In this section we explore model selection by applying existing penalization methods to our proxy-likelihood. 
Specifically, we use the composite likelihood AIC suggested by \citep{varin_vidoni2005note, varin2008composite} and used in \citet{padoan_ribatet_sisson2010}. 
We explain the composite likelihood AIC (CLAIC) and implement it in our proxy-likelihood in Appendix~\ref{ap:model_selection}. 
Targeting the TPD (rather than, e.g., the ability to distinguish between similar models) makes model selection a challenging task. We want a method to perform model selection because we want to know which model best recreates the observed TPD, not because we think we can find the true model.
Our method was not developed to be able to distinguish between two models with similar TPD functions (e.g., a TL-AR(1) and a TL-ARMA(1,1) with small $\theta$).

\begin{figure}
    \centering
    \includegraphics[width=0.7\linewidth]{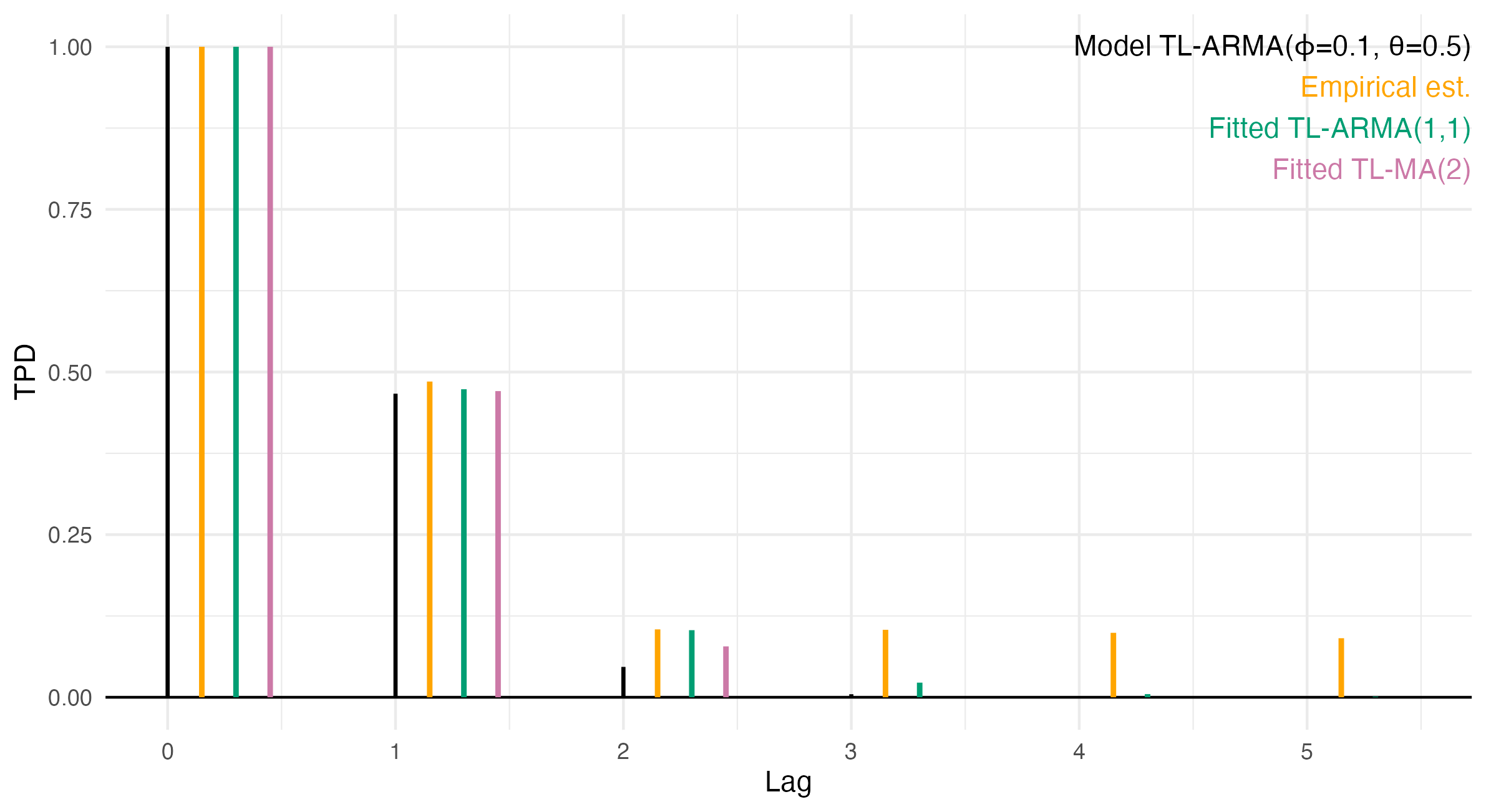}
    \caption{Model TPD in black, empirically estimated TPD from a time series of length 10000 in yellow, and TPD from fitted models. The fitted models are an TL-ARMA(1,1) in green, and TL-MA(2) in pink.}
    \label{fig:arma_ma2_comparison}
\end{figure}

Figure (\ref{fig:arma_ma2_comparison}) shows how similar two fitted models can be and demonstrates that both models fit the data quite well. 
In this case both models (a TL-ARMA(1,1) and a TL-MA(2)) have two parameters though the TL-ARMA is in some sense, when represented as a TL-MA($\infty$), much more complex. 
In this case, the TL-ARMA model fit best when using CLAIC (152771.7 compared to 152776.9). % and CLBIC (by 4.9 units). 
Repeating this experiment, we find that the CLAIC selects the TL-ARMA over the TL-MA(2) model in 35 out of 100 simulated time series. 
In general the scores are very close, 78 of the simulated series have scores within 2 units of each other and 89 have scores within 5 units of each other. 
Our method will not be able to reliably determine that these data came from a TL-ARMA model but it will find which parameters from each model best capture the TPD and will select a model that best captures the TPD information present in the proxy-likelihood.

\begin{figure}
    \centering
    \includegraphics[width=0.95\linewidth]{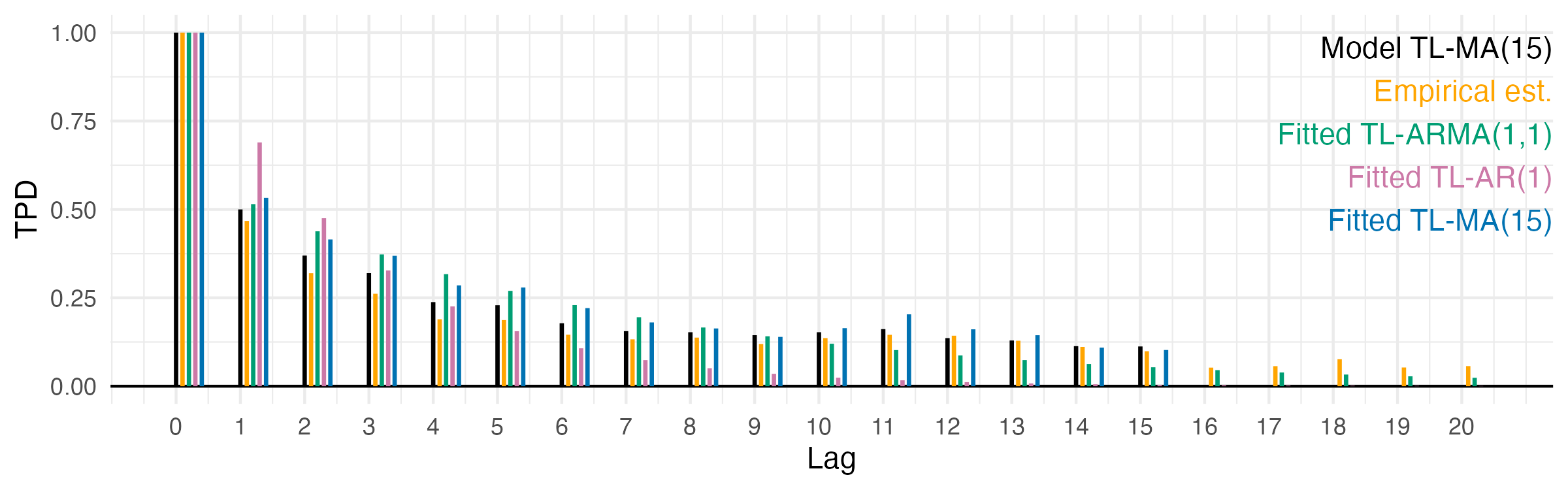}
    \caption{Model TPD in black, empirical TPD from a time series of length 10000 generated from the TL-MA(15) fitted in \citet{wixson_cooley2023attribution} in yellow, and TPD from fitted models. The fitted models are an TL-ARMA(1,1) in green, a TL-AR(1) in pink, a TL-MA(15) (the generating model) in blue, and a TL-MA(20) in grey.}
    \label{fig:ma15_sim_comparison}
\end{figure}

Our final simulation generates data from a TL-MA(15) with the coefficients set to the fitted values obtained from the innovations fit to present climate ERA5-derived FWI data in Grand Lake, Colorado. 
This is the fitted model from \citet[Section 3]{wixson_cooley2023attribution}.
We selected this model as it is an example of a believable TPD from environmental data which does not decay as smoothly as transformed-linear models with some AR component. 
In this simulation study we fit TL-MA models of orders $q = 1, \dots, 20$, the TL-AR(1) model, and the TL-ARMA(1,1) model. 
Figure \ref{fig:ma15_sim_comparison} displays the model, empirical, and fitted TPDs from one simulation. 
We highlight the broad agreement between the TL-ARMA and TL-MA models. 

We consider two new metrics in this simulation study. 
Motivated by common practice in applying AIC, we allow for the selection of a more parsimonious model (the TL-MA(15)) than the best model if it is within two units and term this objective CLAIC-2. 
In addition, we consider standard AIC which penalizes our objective function by two units for each parameter in the model. 
\citet{akaike1974new} gives theoretical justification for this penalty but this justification requires the correct likelihood.

In this set of simulations the generating TL-MA(15) model was chosen a majority of times with all metrics; 75 times with CLAIC, 90 times with CLAIC-2, % 85 times with CLBIC, 
and 93 times with AIC.
Table \ref{tab:ma15_results_table} shows the number of times that each fitted model is selected by the three criterion. 
The selected models were a TL-MA model of order 15 or greater when the generating model was not chosen. %  (with the exception of CLBIC which selected a TL-MA(13) once). 
Each of these selected models can capture the longer-range dependence of the data which has non-zero TPD values out to lag-15.

\begin{table}
\centering

\begin{tabular}{lrrr}
  \hline
  Model 		& CLAIC 	& CLAIC-2  	&AIC \\ 		% & CLBIC
  \hline
  TL-AR(1) 	&   0 		&   0 			& 0\\ 			% &   0 	
  TL-MA(1) 	&   0 		&   0 		 	& 0\\ 			% &   0
   \vdots 		&   \vdots 	& \vdots 	 	&  \vdots \\ 	% & \vdots
  TL-MA(12) 	&   0 		&   0 			& 0 \\ 		% &  0 
  TL-MA(13) 	&   0 		&   0 		 	& 0 \\ 		% &  1
  TL-MA(14) 	&   0 		&   0 		 	&  0 \\ 		% & 0
  TL-MA(15) 	&  {\bf 75} 	&  {\bf 90}  	&  {\bf 93} \\	% & {\bf 85}
  TL-MA(16) 	&  5 		&  0 			&  2 \\ 		% & 4 
  TL-MA(17) 	&  4 		&  0 			& 1\\ 			% &  0 
  TL-MA(18) 	&  4 		&   1 			& 1 \\ 		% &  2 	
  TL-MA(19) 	&  3 		&   1 		 	& 1 \\ 		% &  2
  TL-MA(20) 	&  9 		&   8 		 	& 3 \\ 		 % &  6
  TL-ARMA(1,1) & 0 		&  0 			& 0 \\ 		% &  0 	
   \hline
\end{tabular}
\caption{Number of times (out of 100 simulations) that each model is selected by CLAIC, CLAIC which allows selection of MA(15) if it is within 2 units of the best model, and the penalty from basic AIC, when data are generated from TL-MA(15). Bold indicates the model with the most selections from that criterion.} 
\label{tab:ma15_results_table}
\end{table}

\section{Case Study: Wildfire Data}\label{sec:case_study}
We apply our method to the FWI data from Grand Lake, Colorado used to study wildfire risk in \citet{wixson_cooley2023attribution}. 
In that paper, the authors compute the daily FWI from ERA5 weather variables for June 1 through October 31 under past climate (1958-1979) and present climate (2002-2021). 
These two sets of seasonal wildfire risk time series are assumed to have come from distinct processes which the authors model by estimating the seasonality, performing a marginal transformation, and then fitting TL-MA models to the transformed time series to capture the dependence in the upper tail of the processes. 
Model selection was performed in an \textit{ad hoc} manner and models with an AR component were not considered.  
A notable marginal shift in the upper tail was found to exist for the different time periods but there was not much difference in the dependence structure.

Previous work with these TLETS models \citep{wixson_cooley2023attribution, mhatre_cooley2024} include a bias reduction step to mitigate known bias in the empirical TPD estimator \citep{huser_davison_genton_2016}. 
That bias reduction involved subtracting the mean of the time series from each value and setting negative observations to zero prior to TPD estimation. 
When considering pairwise dependence, shifting the time series towards zero has the intended effect of shifting points that are near the axes toward (or onto) the axes without much change to the angles of points that lie near the identity line. 
This reduces the estimated dependence at large lags (when true dependence is negligible) without much change to the lags with strong dependence.
While reducing bias in the later lags, this step introduces bias in TPD estimates by changing the angles for all points not on the identity line. 
These angular changes are largest for points that are closest to the axes with smaller radial components and thus lags with moderate to weak model dependence see a negative bias in the estimated dependence. 
Our proxy-likelihood method does not require a bias-reduction step. 
In fact, this shift toward the axes changes the marginal distribution which would result in our proxy-likelihood picking up on marginal mis-match while also trying to capture the angle-adjusted dependence structure. 
Thus, the plots in this section do not have the bias-reduction and therefore do not match plots in the previous study.

As in the previous simulation study, we fit TL-MA models of orders $q = 1, \dots, 20$, the TL-AR(1) model, and the TL-ARMA(1,1) model. 
We use the 0.95 quantile of the radial components as the censoring threshold and use 20 sub-series to estimate the covariance of the score (needed for the CLAIC penalty, Appendix~\ref{ap:model_selection}). 
Each sub-series corresponds to a single wildfire season and thus treating them as replicates matches what was done in \citet{wixson_cooley2023attribution} and is justifiable beyond being necessary for estimation of the covariance of the score. 
We expect that many environmental applications will have natural sub-series that can be exploited similarly.

\begin{figure}
    \centering
    \includegraphics[width=0.95\textwidth]{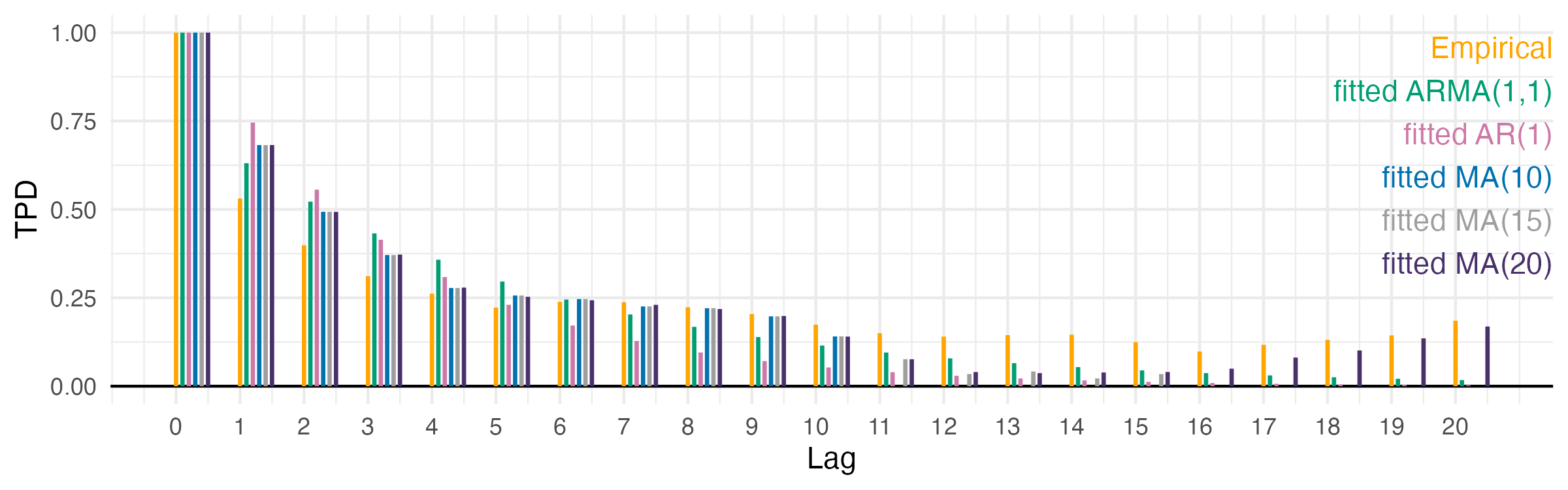}
    \caption{Tail Pairwise Dependence plot from past climate ERA5 FWI in Colorado. In yellow is the empirically estimated TPD (\ref{eq:tpdf_empirical}). Each color is the model-based TPD from the respectively fitted model.}
    \label{fig:early_wildfire}
\end{figure}

Our method captures the second-order behavior in the past climate (Figure \ref{fig:early_wildfire}). 
There is strong agreement between the fitted TPD values of the TL-ARMA model and the TL-MA models for many lags. 
This highlights that our method can be considered a TPD estimator which differs from the empirical estimator as was seen in Section~\ref{sec:tpd_estimation}. 
The proxy-likelihood suggests that there is stronger dependence at short lags and weaker dependence at later lags than what is indicated by the empirical estimator. 
This difference between the empirical dependence estimates and the dependence estimated by our models is likely bias as discussed previously.
It may also include bias due to an angular measure mismatch between the true data generating process, the HR likelihood, and the TLETS models that we are fitting.
The TL-AR and TL-ARMA fits illustrate that simpler models handle some of this mismatch by balancing the fit between early and late lags. 
Here that means that the fitted TL-AR overestimates dependence at lag-1 and underestimates dependence in lag-6 and beyond whereas the TL-ARMA appears to underestimate dependence at lag-1 and overestimate at lag-4. 

Table \ref{tab:early_proxy_fitted_info} gives an overview of the model selection information that is readily obtained from our proxy-likelihood method. 
Values in the table columns entitled ``$-\ell_{cl}(\hat{\btheta}_{MPL})$'', ``CLAIC'', and ``Basic AIC'' are the difference between the value of the objective function for that model and the lowest (best) value of the objective function. 
The negative log-likelihood evaluated at the maximum proxy-likelihood estimate, $-\ell_{cl}(\hat{\btheta}_{MPL})$, shows the expected behavior; more complex models have lower deviation from the best fitting model. 
In other words, as the order of the TL-MA increases, the value of the negative composite log-likelihood decreases. 
It is no surprise that the most complicated model, the TL-MA(20), has the lowest (best) score.
Our proxy-likelihood suggests that the TL-AR model is a better fit than a TL-MA(5) which is not surprising as the TPD plot indicates non-negligible dependence beyond lag-5. 

\begin{table}
\centering
\begin{tabular}{l|r|rr|rr}
  Model & $-\ell_{cl}(\hat{\boldsymbol\theta}_{MPL})$ & CL Penalty & CLAIC & Params & Basic AIC \\ 
  \hline
  TL-AR(1) 	 & 45.53   & 2.12	& 87.98   	&   1 		& 53.05 \\ 
  TL-MA(1) 	 & 278.62	& 0.00    	& 549.93	&   1 		& 519.25 \\ 
  TL-MA(2) 	 & 183.20 & 0.01  	& 359.11	&   2 		& 330.41  \\ 
  TL-MA(3) 	 & 132.23 	& 0.04  	& 257.21	&   3 		& 230.45  \\ 
  TL-MA(4) 	 & 107.11  & 0.71  	& 208.33	&   4 		& 182.22  \\ 
  TL-MA(5) 	 & 87.91   & 0.92  	& 170.35	&   5 		& 145.83  \\ 
  TL-MA(10)	 & 23.59   & 2.00  	& 43.85	&  10 	& 27.17  \\ 
  TL-MA(15) 	 & 21.44   & 2.89  	& 41.34	&  15 	& 32.89 \\ 
  TL-MA(20) 	 & 0     	& 3.66  	& 0		&  20		& 0  \\ 
  TL-ARMA(1,1) & 24.16 	& 2.84 	& 46.67	&   2 		& 12.31
\end{tabular}
\caption{Information from proxy-likelihood fitted models for Past Climate FWI time series from ERA5 data in Colorado. Scores are listed as difference from the the best score where ``0'' indicates the model with the best score. The second column is the negative log-likelihood evaluated at the maximum proxy-likelihood estimate, the CL Penalty is in Appendix~\ref{ap:model_selection}, Params indicates the number of parameters, and Basic AIC penalizes with 2$\times$Params.}
\label{tab:early_proxy_fitted_info}
\end{table}

The CLAIC penalties are increasing with model complexity (Table \ref{tab:early_proxy_fitted_info}, column 3, labeled ``CL Penalty''). 
These penalties are very small; the penalty applied to the TL-MA(20) is less than twice that of the TL-ARMA model. 
This observation led us to consider penalizing the likelihood with the penalty from classical AIC \citep{akaike1974new}. 
Both penalization methods select a TL-MA(20) as the best fitting model which is plausible given the observed long dependence and complex shape of the estimated TPD. 
The two penalties result in different ordering of the models; the CLAIC ranks the TL-ARMA(1,1) as worse than a TL-MA(10) while the AIC ranks the TL-ARMA(1,1) as the second best model. 
The apparent increase in dependence at lag-20 is something that TL-MA models of large enough order can capture but a TL-ARMA(1,1) model cannot capture. 
As a check, we refit the models considering dependence out to lag-19 which results in AIC choosing the TL-ARMA(1,1).

\begin{figure}
    \centering
    \includegraphics[width=0.95\textwidth]{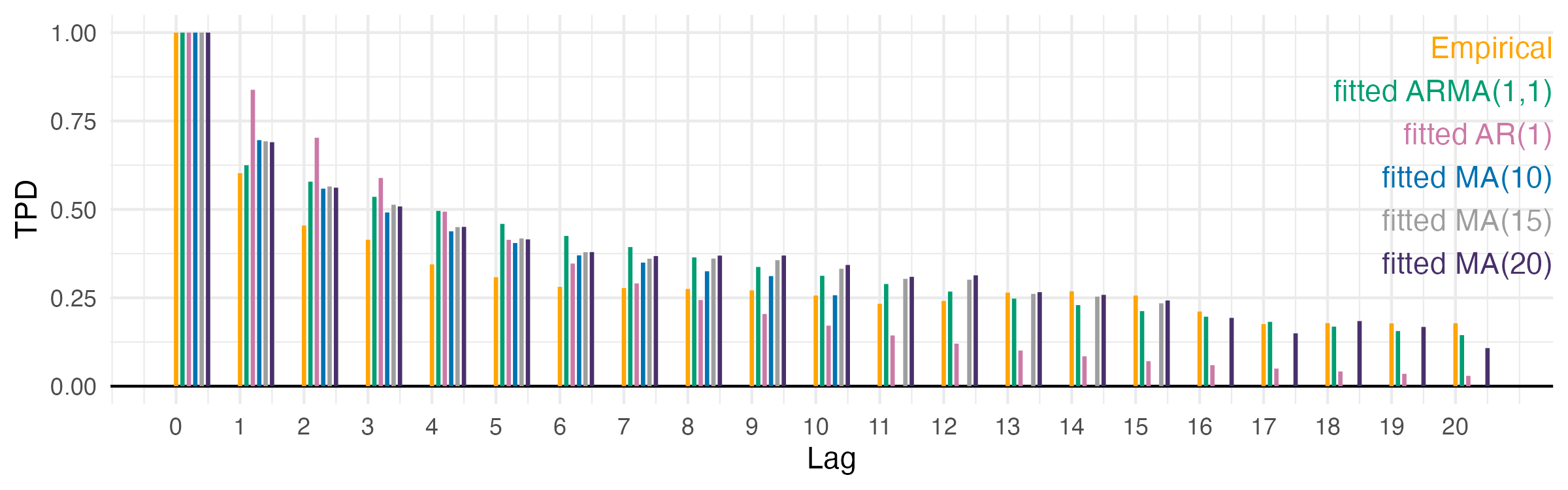}
    \caption{Tail Pairwise Dependence plot from present climate ERA5 FWI in Colorado (as in Figure \ref{fig:early_wildfire})}
    \label{fig:late_wildfire}
\end{figure}

The estimated TPD from the present climate data is plotted in Figure \ref{fig:late_wildfire}. 
This TPD plot displays stronger and longer dependence than what was observed in the past climate (Figure~\ref{fig:early_wildfire}). 
This apparent change in dependence was obscured by the bias correction in \citet{wixson_cooley2023attribution}. 
We also note that our estimates of dependence are larger than the dependence estimated by the empirical estimator which we expect to be due to the included radial information.

Here, as with the past climate, the CLAIC selects the TL-MA(20) but the AIC penalty selects the TL-ARMA(1,1) (Table~\ref{tab:late_proxy_fitted_info}). 
While both models capture the character of dependence, the TL-ARMA(1,1) appears to smooth some of the observed complex shape of the TPD function. 
We expect many practitioners would be happy to choose the model with 13 fewer parameters though further study is needed to determine which penalty is better.
In the past climate results (Table~\ref{tab:early_proxy_fitted_info}) we see a leveling off of the CLAIC and AIC in models that can capture dependence out to lag-10, here that leveling occurs around lag-15, this echos the longer range dependence observed in the TPD plots. 

\begin{table}
\centering
\begin{tabular}{l|r|rr|rr}
  Model & $-\ell_{cl}(\hat{\boldsymbol\theta}_{MPL})$ & CL Penalty & CLAIC & Params & Basic AIC \\ 
  \hline
  TL-AR(1) 	 & 143.99  & 2.84	& 288.66	&   1 		& 265.52 \\ 
  TL-MA(1) 	 & 631.60	& 0.00    	& 1258.20	&   1 		& 1240.75 \\ 
  TL-MA(2) 	 & 521.45 & 0.03  	& 1037.95	&   2 		& 1022.44  \\ 
  TL-MA(3) 	 & 434.67 	& 0.07  	& 864.47	&   3 		& 850.88  \\ 
  TL-MA(4) 	 & 367.60  & 0.12  	& 730.45	&   4 		& 718.75  \\ 
  TL-MA(5) 	 & 313.91  & 0.16 	& 623.13	&   5 		& 613.36  \\ 
  TL-MA(10)	 & 132.50  & 0.38  	& 260.76	&  10 	& 260.55  \\ 
  TL-MA(15) 	 & 26.36   & 0.92  	& 49.56	&  15 	& 58.27 \\ 
  TL-MA(20) 	 & 0     	& 2.50  	& 0		&  20		& 15.55  \\ 
  TL-ARMA(1,1) & 10.23 	& 1.61 	& 18.67	&   2 		& 0
\end{tabular}
\caption{Information from proxy-likelihood fitted models for Present Climate FWI time series from ERA5 data in Colorado as in Table \ref{tab:early_proxy_fitted_info}. Scores are listed as difference from the the best score thus ``0'' is the best score.}
\label{tab:late_proxy_fitted_info}
\end{table}

\section{Discussion}
\label{sec:proxy_discussion}

We have developed a maximum composite proxy-likelihood estimator for regularly varying models that have intractable likelihoods and applied it to the TLETS models of \citet{mhatre_cooley2024}. 
While developed in the context of these TLETS models, adapting the method to other models is straightforward as all that is needed is the map between the parameters of the target model and the TPD (the analogues of (\ref{eq:AR1_TPD}), (\ref{eq:MAq_TPD}), and (\ref{eq:ARMA11_TPD})). 
Our method is able to capture the second-order tail behavior but, like all methods for the tail, is affected by the challenges inherent to estimating an asymptotic quantity (the TPD) at finite levels. 
The linking of two models through a summary measure implies an angular measure mismatch. 
We expect this feature to allow for better fitting to real data that do not have discrete angular measures but the mathematical implications are not well understood. 

Simulations suggest that the proxy-likelihood outperforms existing estimators. 
The proxy-likelihood based TPD estimator outperforms the empirical TPD estimator when the dependence is weak which we attribute to the retained radial information.
TLETS model fitting is improved by the proxy-likelihood when compared to existing methods. 
Likelihood-based model selection methods are straightforward to implement and provide a principled method for selecting from the library of TLETS models. 

Application to the wildfire risk data used in \citet{wixson_cooley2023attribution} highlights the benefits of our proxy-likelihood. 
This method allows us to see that the bias reduction in previous work obscured a shift in the tail dependence structure between past and present climate. 
In addition, the principled model selection methods that follow from having a likelihood-like method provide information about the benefits of using more complex models. 

The composite likelihood penalties in section~\ref{sec:case_study} are quite small which may indicate that they are not well calibrated and that using basic AIC penalties may be preferable. 
The CLAIC penalty requires the estimation of the covariance of the score which is known to be challenging. 
The value of this work is to provide a likelihood-like function which can be readily used with standard model selection methods like penalization or cross-validation.  Identification and justification of an optimal model selection approach is an area of future investigation. 

While implementation of the method is intuitively simple, optimization can be challenging. 
We have found that optimization routines which require the user to specify initial values are somewhat sensitive to the initial values in that they fail at some values and converge at others. 
In practice we have found the most success with initializing all parameters to 0.1, testing for convergence, randomly generating initial values if convergence failed, and repeating if necessary.

\section*{Data availability statement}
Data and scripts for our method and for simulations performed in this manuscript are available on \url{https://github.com/twixson/proxy-lhood}. 
References for data used in Section~\ref{sec:case_study} are included. 
All analyses were performed using R \citep{R}. 
We used the following packages: \texttt{evd} \citep{evd}, \texttt{Matrix} \citep{Matrix}, \texttt{tidyverse} \citep{tidyverse}, and \texttt{cowplot} \citep{cowplot}. 

\section*{Funding statement}
TW and DC were partially supported by US National Science Foundation grant DMS-2311164

\section*{Conflict of interest disclosure}
The authors are not aware of conflicts of interest. 

\section*{Ethical statements}
Not Applicable

\newpage

\bibliographystyle{apalike}
\bibliography{2references}  

\newpage

\appendix
\section{Appendix 1: The HR model}\label{ap:hr_model}

The TLETS models are defined on regularly varying $\alpha = 2$ margins (e.g., Frechet($2$)) so we transform the margins of the HR distribution (\ref{eq:HR_df_Gumbel}) to be Frechet($2$). 
For notational convenience let $a_{ij} = a (x_i, x_j, \lambda_{ij}) = \lambda_{ij} - 2\log(x_i / x_j) / \lambda_{ij}$ and $a_{ji} = a (x_j, x_i, \lambda_{ij}) = \lambda_{ij} - 2\log(x_j / x_i) / \lambda_{ij}$. 
Let $G_{\lambda_{ij}}(x_i, x_j) = F_{\lambda_{ij}}(\log(x_i^2), \log(x_j^2))$ which has Frechet(2) margins and is given by
% Frechet(2) HR distribution
\begin{align}
    G_{\lambda_{ij}}(x_i, x_j) = & \exp \left\{ - x_i^{-2} \Phi \left( a_{ij} \right) - x_j^{-2} \Phi \left( a_{ji} \right) \right\} 
    \notag\\
    = & \exp \left\{ -V(x_i, x_j, \lambda_{ij}) \right\}.
\end{align}
Hereafter we suppress the arguments for $V$. The density is 
% density of Frechet(2) HR dist
\begin{align}\label{eq:HR_density_lambdas_app}
    g_{\lambda_{ij}}(x_i, x_j) =& \frac{\partial^2}{\partial x_i \partial x_j} G_{\lambda_{ij}}(x_i, x_j)
    \notag\\
    = & G_{\lambda_{ij}}(x_i, x_j) \bigg( \frac{\partial}{\partial x_i} V \frac{\partial}{\partial x_j} V - \frac{\partial^2}{\partial x_i \partial x_j} V \bigg)
\end{align}
where the derivatives in (\ref{eq:HR_density_lambdas_app}) are
\begin{align}\label{eq:density_derivs_1_app}
    \frac{\partial}{\partial x_i} V 
    = & \frac{-2}{x_i^3} \Phi (a_{ij}) - \frac{1}{\lambda_{ij} x_i^3} \phi (a_{ij}) + \frac{1}{\lambda_{ij} x_i x_j^2} \phi (a_{ji}) 
    \\\label{eq:density_derivs_2}
    \frac{\partial^2}{\partial x_i \partial x_j} V 
    = & \frac{-\lambda_{ij}^2 - \log\frac{x_i}{x_j}}{\lambda_{ij}^3 x_i^3 x_j} \phi (a_{ij}) + \frac{-\lambda_{ij}^2 - \log\frac{x_j}{x_i}}{\lambda_{ij}^3 x_i x_j^3} \phi (a_{ji}).
\end{align} 

\newpage 

\section{Appendix 2: TPD link}\label{ap:tpd_link}
Following Theorem 1 in \citet{coles_tawn1991modelling}, we note that we can obtain the angular density $h^*_{\lambda_{ij}}(\cdot)$ of the HR model by taking partial derivatives of the exponent measure $V$ and transforming to pseudo-polar coordinates. 
As mentioned above, it is convenient for the models that we will work with to use the $L_2$-norm and for the margins to be Frechet(2). 
This means that we cannot simply apply Theorem 1 from Coles and Tawn. In their case the Jacobian of the transformation is the $L_1$-norm ($x_i + x_j$) whereas ours includes an extra term. 
Let $r = ||(x_i, x_j)||_2 = \sqrt{x_i^2 + x_j^2}$ and $\mathbf{s} = (s_i, s_j) = (x_i, x_j)/r$ so that $r$ is the radial component of each point and $\mathbf{s}$ is the pseudo-angular component. Lemma 1.1 in \citet{song_gupta1997} show that the Jacobian is $r/s_j$. 
We have already taken the derivatives in (\ref{eq:HR_density_lambdas}) which we rewrite in anticipation of the pseudo-polar change of variables:
\begin{align}
    \frac{\partial^2}{\partial x_i \partial x_j} V = & \mu(d\mathbf{x}_{ij})
    \notag\\
    = & \Bigg\{\frac{-\lambda_{ij}^2 - \log\frac{x_i}{x_j}}{\lambda_{ij}^3 x_i^3 x_j} \phi (a_{ij}) + \frac{-\lambda_{ij}^2 - \log\frac{x_j}{x_i}}{\lambda_{ij}^3 x_i x_j^3} \phi (a_{ji}) \Bigg\} d\mathbf{x}_{ij}
    \notag\\
    = & \Bigg\{\left(x_i^2 + x_j^2\right)^{-2} \frac{-\lambda_{ij}^2 - \log\frac{x_i/\sqrt{x_i^2 + x_j^2}}{x_j/\sqrt{x_i^2 + x_j^2}}}{\lambda_{ij}^3 x_i^3 x_j \left(x_i^2 + x_j^2\right)^{-2}} \phi \left(\lambda_{ij} - \frac{1}{\lambda_{ij}} \log \frac{\frac{x_i}{\sqrt{x_i^2 + x_j^2}}}{\frac{x_j}{\sqrt{x_i^2 + x_j^2}}}\right) +
    \notag\\
    & \hspace{0.1in} \left(x_i^2 + x_j^2\right)^{-2} \frac{-\lambda_{ij}^2 - \log\frac{x_j/\sqrt{x_i^2 + x_j^2}}{x_i/\sqrt{x_i^2 + x_j^2}}}{\lambda_{ij}^3 x_i x_j^3 \left(x_i^2 + x_j^2\right)^{-2}} \phi \left(\lambda_{ij} - \frac{1}{\lambda_{ij}} \log \frac{\frac{x_j}{\sqrt{x_i^2 + x_j^2}}}{\frac{x_i}{\sqrt{x_i^2 + x_j^2}}}\right) \Bigg\} d\mathbf{x}_{ij}. 
\end{align}
To transform to polar coordinates we include the Jacobian
\begin{align}
    \mu(d\mathbf{x}_{ij}) = & J(\mathbf{x}_{ij} \rightarrow r, \mathbf{s}) \mu(dr \times d\mathbf{s}) 
    \notag\\
    = & \frac{r}{s_j} r^{-4} \Bigg\{ \frac{- \lambda_{ij}^2 - \log\frac{s_i}{s_j}}{\lambda_{ij}^3 s_i^3 s_j} \phi \left(\lambda_{ij} - \frac{1}{\lambda_{ij}} \log \frac{s_i}{s_j}\right) +
    \notag\\
    & \hspace{0.25in} \frac{-\lambda_{ij}^2 - \log\frac{s_j}{s_i}}{\lambda_{ij}^3 s_i s_j^3 } \phi \left(\lambda_{ij} - \frac{1}{\lambda_{ij}} \log \frac{s_j}{s_i}\right) \Bigg\} dr d\mathbf{s}
    \notag\\
    = & -2 r^{-3} dr \hspace{0.05in} \frac{1}{2s_j}\Bigg\{ \frac{\lambda_{ij}^2 + \log\frac{s_i}{s_j}}{\lambda_{ij}^3 s_i^3 s_j} \phi \left(\lambda_{ij} - \frac{1}{\lambda_{ij}} \log \frac{s_i}{s_j}\right) +
    \notag\\
    & \hspace{0.25in} \frac{\lambda_{ij}^2 + \log\frac{s_j}{s_i}}{\lambda_{ij}^3 s_i s_j^3 } \phi \left(\lambda_{ij} - \frac{1}{\lambda_{ij}} \log \frac{s_j}{s_i}\right) \Bigg\}  d\mathbf{s}
    \notag\\
    = & -2 r^{-3} dr \hspace{0.05in} h_{\lambda_{ij}}(\mathbf{s})d\mathbf{s}
\end{align}
Here we note that $h_{\lambda_{ij}}(\mathbf{s})$ is a density on the one dimensional $L_2$-ball in the positive quadrant of $\mathbb{R}^2$ and thus it is equivalent to 
\begin{align}
    h_{\lambda_{ij}}(s_i) = & \frac{1}{2 \sqrt{1-s_i^2}}\Bigg\{ \frac{\lambda_{ij}^2 + \log\frac{s_i}{\sqrt{1-s_i^2}}}{\lambda_{ij}^3 s_i^3 \sqrt{1-s_i^2}} \phi \left(\lambda_{ij} - \frac{1}{\lambda_{ij}} \log \frac{s_i}{\sqrt{1-s_i^2}}\right) +
    \notag\\
    & \hspace{0.25in} \frac{\lambda_{ij}^2 + \log\frac{\sqrt{1-s_i^2}}{s_i}}{\lambda_{ij}^3 s_i \sqrt{1-s_i^2}^3 } \phi \left(\lambda_{ij} - \frac{1}{\lambda_{ij}} \log \frac{\sqrt{1-s_i^2}}{s_i}\right) \Bigg\}. 
\end{align}

\newpage 

\subsection{Map between HR parameter and TPD}\label{ap:hr_tpd}
The $ij^{th}$ parameter of the HR distribution has $ij^{th}$ TPD parameter
\begin{align}\label{eq:HR_TPD_app}
    \sigma_{ij} = & \int_0^1 s_i \sqrt{1-s_i^2} h_{\lambda_{ij}}(s_i) ds_i 
    \notag\\ 
    = & \frac{1}{2} \int_0^1 s_i \Bigg\{ \frac{\lambda_{ij}^2 + \log\frac{s_i}{\sqrt{1-s_i^2}}}{\lambda_{ij}^3 s_i^3 \sqrt{1-s_i^2}} \phi \left(\lambda_{ij} - \frac{1}{\lambda_{ij}} \log \frac{s_i}{\sqrt{1-s_i^2}}\right)
    \notag\\
    & \hspace{1.5in} + \frac{\lambda_{ij}^2 + \log\frac{\sqrt{1-s_i^2}}{s_i}}{\lambda_{ij}^3 s_i \sqrt{1-s_i^2}^3 } \phi \left(\lambda_{ij} - \frac{1}{\lambda_{ij}} \log \frac{\sqrt{1-s_i^2}}{s_i}\right) \Bigg\} ds_i \\
  \text{Perform a } u-&\text{substitution with }\hspace{0.2in} u = \log\frac{s_i}{\sqrt{1-s_i^2}} \\
  % \implies \exp{2u} = \frac{s_i^2}{1-s_i^2} 
  % \implies & s_i^2 = \exp{2u} - s_i^2 \exp{2u} 
  % \implies s_i^2 =  \frac{\exp{2u}}{1 + \exp{2u}}\\
  \implies s_i = & \frac{\exp{u}}{\sqrt{1 + \exp{2u}}}, \hspace{0.2in}
  \implies \sqrt{1-s_i^2} = \frac{1}{\sqrt{1 + \exp{2u}}}, \hspace{0.2in}
  \implies ds_i = \frac{\exp{u}}{(1 + \exp{2u})^{\frac{3}{2}}} du \\
\text{It is easy to see}&\text{ that the bounds for integration change to } (-\infty, \infty) \\
    % \sigma_{ij} = & \frac{1}{2 \lambda_{ij}^3} \int_0^1 \Bigg\{ \frac{\lambda_{ij}^2 + \log\frac{s_i}{\sqrt{1-s_i^2}}}{ s_i^2 \sqrt{1-s_i^2}} \phi \left(\lambda_{ij} - \frac{ \log \frac{s_i}{\sqrt{1-s_i^2}}}{\lambda_{ij}}\right) +
    % \frac{\lambda_{ij}^2 + \log\frac{\sqrt{1-s_i^2}}{s_i}}{ \sqrt{1-s_i^2}^3 } \phi \left(\lambda_{ij} - \frac{\log \frac{\sqrt{1-s_i^2}}{s_i}}{\lambda_{ij}} \right) \Bigg\} ds_i \\ 
    % = & \frac{1}{2 \lambda_{ij}^3} \int_{-\infty}^\infty \frac{\exp{u}}{(1 + \exp{2u})^{\frac{3}{2}}}  \Bigg\{ \frac{\lambda_{ij}^2 + u}{ \frac{\exp{2u}}{1 + \exp{2u}}  \frac{1}{\sqrt{1 + \exp{2u}}} }  \phi \left(\lambda_{ij} - \frac{u}{\lambda_{ij}}\right) + \frac{\lambda_{ij}^2 - u}{  \frac{1}{(1 + \exp{2u})^{\frac{3}{2}}} } \phi \left(\lambda_{ij} + \frac{u}{\lambda_{ij}} \right) \Bigg\} du \\
    = & \frac{1}{2 \lambda_{ij}^3} \int_{-\infty}^\infty  \Bigg\{ \frac{\exp{u}}{(1 + \exp{2u})^{\frac{3}{2}}}  \frac{\lambda_{ij}^2 + u}{ \frac{\exp{2u}}{1 + \exp{2u}}  \frac{1}{\sqrt{1 + \exp{2u}}} }  \phi \left(\lambda_{ij} - \frac{u}{\lambda_{ij}}\right)   \\
    & \hspace{1.5in} + \frac{\exp{u}}{(1 + \exp{2u})^{\frac{3}{2}}}\frac{\lambda_{ij}^2 - u}{  \frac{1}{(1 + \exp{2u})^{\frac{3}{2}}} } \phi \left(\lambda_{ij} + \frac{u}{\lambda_{ij}} \right)  \Bigg\} du \\ 
    % = &  \frac{1}{2 \lambda_{ij}^3}  \int_{-\infty}^\infty \Bigg\{ \frac{\lambda_{ij}^2 + u}{ \exp{u} }  \phi \left(\lambda_{ij} - \frac{u}{\lambda_{ij}}\right)  +  \exp{u}(\lambda_{ij}^2 - u) \phi \left(\lambda_{ij} + \frac{u}{\lambda_{ij}} \right)  \Bigg\} du  \\ 
    = & \frac{1}{2 \lambda_{ij}^3}  \int_{-\infty}^\infty \Bigg\{ \frac{\lambda_{ij}^2 + u}{ \exp{u} } (2 \pi)^{-\frac{1}{2}} \exp \left[ -\frac{1}{2} \left(\lambda_{ij} - \frac{u}{\lambda_{ij}}\right)^2 \right]  \\
    & \hspace{1.5in} +  \exp{u}(\lambda_{ij}^2 - u) (2 \pi)^{-\frac{1}{2}} \exp \left[ -\frac{1}{2} \left(\lambda_{ij} + \frac{u}{\lambda_{ij}}\right)^2 \right]  \Bigg\} du  \\ 
    % = & \frac{1}{2 (2 \pi)^{\frac{1}{2}} \lambda_{ij}^3}  \int_{-\infty}^\infty \Bigg\{ (\lambda_{ij}^2 + u)  \exp \left[ - u -\frac{1}{2} \left(\lambda_{ij}^2 - 2u +  \frac{u^2}{\lambda_{ij}^2}\right) \right]  \\
    % & \hspace{1.5in} +  (\lambda_{ij}^2 - u) \exp \left[ u -\frac{1}{2} \left(\lambda_{ij}^2 + 2u + \frac{u^2}{\lambda_{ij}^2}\right) \right]  \Bigg\} du  \\ 
    = & \frac{1}{2 (2 \pi)^{\frac{1}{2}} \lambda_{ij}^3}  \int_{-\infty}^\infty \Bigg\{ (\lambda_{ij}^2 + u) \exp \left( - u - \frac{\lambda_{ij}^2}{2}  + u -  \frac{u^2}{2\lambda_{ij}^2} \right)  \\
    & \hspace{1.5in} +  (\lambda_{ij}^2 - u)  \exp \left( u -\frac{\lambda_{ij}^2}{2} - u - \frac{u^2}{2\lambda_{ij}^2} \right)  \Bigg\} du  \\ 
    % = & \frac{1}{2 (2 \pi)^{\frac{1}{2}} \lambda_{ij}^3}  \int_{-\infty}^\infty \Bigg\{ (\lambda_{ij}^2 + u) \exp \left( - \frac{\lambda_{ij}^2}{2} -  \frac{u^2}{2\lambda_{ij}^2} \right) +  (\lambda_{ij}^2 - u) \exp \left( -\frac{\lambda_{ij}^2}{2} - \frac{u^2}{2\lambda_{ij}^2} \right)  \Bigg\} du  \\ 
    % = & \frac{1}{2 (2 \pi)^{\frac{1}{2}} \lambda_{ij}^3}  \int_{-\infty}^\infty 2\lambda_{ij}^2 \exp \left( - \frac{\lambda_{ij}^2}{2} -  \frac{u^2}{2\lambda_{ij}^2} \right) du  \\ 
    = & \frac{1}{ (2 \pi)^{\frac{1}{2}} \lambda_{ij}}  \int_{-\infty}^\infty \exp \left( - \frac{\lambda_{ij}^2}{2} -  \frac{u^2}{2\lambda_{ij}^2} \right) du  \\ 
    \text{Another } u-&\text{substitution with }\hspace{0.2in} v = \frac{u}{\lambda_{ij}}  \implies du = \lambda_{ij} dv \\ 
    = & \frac{1}{(2 \pi)^{\frac{1}{2}} \lambda_{ij}} \int_{-\infty}^\infty \lambda_{ij} \exp \left( - \frac{\lambda_{ij}^2}{2} - \frac{1}{2} v^2 \right) dv  \\ 
    = & \exp \left( - \frac{\lambda_{ij}^2}{2} \right) \int_{-\infty}^\infty  (2\pi)^{\frac{1}{2}} \exp \left( - \frac{1}{2} v^2 \right) dv  \\ 
    \text{Where the } & \text{integrand is a standard normal density and thus} \\
    \sigma_{ij} = & \exp \left( - \frac{\lambda_{ij}^2}{2} \right) \\
    \implies \lambda_{ij} = & \sqrt{-2 \log \sigma_{ij}}
\end{align}

\newpage 

\section{Appendix 3: Score function}\label{ap:score_func}

The contribution of one point to the score function for some $\theta_k \in \btheta$ is, by the chain rule, of the form 
\begin{align}\label{eq:score_contrib_ap}
    \frac{\partial}{\partial \theta_k} \ell_{cl}[\lambda\{\sigma(\btheta)\} | x_n, x_{n+h}] = & \frac{\partial}{\partial \lambda_h} \ell_{cl}(\lambda_h | x_n, x_{n+h}) \frac{\partial}{\partial \sigma} \lambda(\sigma) \frac{\partial}{\partial \theta_k} \sigma(h, \btheta).
\end{align}

Let $a_{h} = a (x_n, x_{n+h}, \lambda_{h}) = \lambda_{h} - \log(x_n / x_{n+h}) / \lambda_{h}$ and let $a_{h'} = a (x_{n+h}, x_n, \lambda_{h})$ be defined by symmetry so that $a_h$ and $a_{h'}$ are analogous to $a_{ij}$ and $a_{ji}$ used above. The first component of the score for one lag-$h$ point is (assuming unit weights) 
\begin{align}\label{eq:component_one}
    \frac{\partial}{\partial \lambda_{h}} \ell_{cl} (\lambda_h | x_n, x_{n+h}) 
    = & 
    - x_n^{-2} \frac{\partial}{\partial \lambda_{h}} & \Phi (a_{h}) - x_{n+h}^{-2} \frac{\partial}{\partial \lambda_{h}} \Phi (a_{h'}) + 
    \frac{\frac{\partial}{\partial \lambda_{h}} \left( \frac{\partial}{\partial x_n} V \frac{\partial}{\partial x_{n+h}} V  - \frac{\partial^2}{\partial x_n \partial x_{n+h}} V \right)}{ \frac{\partial}{\partial x_n} V \frac{\partial}{\partial x_{n+h}} V  - \frac{\partial^2}{\partial x_n \partial x_{n+h}} V }.
\end{align}
The derivatives in (\ref{eq:component_one}) are 
\begin{align}
    \frac{\partial}{\partial \lambda_{h}} \Phi (a_{h}) = & \left(1 + \frac{1}{\lambda_{h}^2} \log \frac{x_n}{x_{n+h}} \right) \phi (a_{h})
    \\
    \frac{\partial}{\partial \lambda_{h}} \left( \frac{\partial}{\partial x_n} V \frac{\partial}{\partial x_{n+h}} V \right) = & \frac{\partial}{\partial x_{n+h}} V \left( \frac{\partial}{\partial \lambda_{h}} \frac{\partial}{\partial x_n} V \right) + \frac{\partial}{\partial x_n} V \left( \frac{\partial}{\partial \lambda_{h}} \frac{\partial}{\partial x_{n+h}} V \right) 
    \notag\\
    \frac{\partial}{\partial \lambda_{h}} \frac{\partial}{\partial x_n} V  
    = & \frac{-2}{x_n^3} \left(1 + \frac{1}{\lambda_{h}^2} \log \frac{x_n}{x_{n+h}} \right) \phi (a_{h}) + 
    \notag\\
    & \hspace{0.25in} \frac{\lambda_{h}^2 + 1 - \frac{1}{\lambda_{h}^4} \log^2 \frac{x_n}{x_{n+h}}}{x_n^3} \phi (a_{h}) - 
    \notag\\ 
    & \hspace{0.25in} \frac{\lambda_{h}^2 + 1 - \frac{1}{\lambda_{h}^4} \log^2 \frac{x_{n+h}}{x_n}}{x_n x_{n+h}^2} \phi (a_{h'})
\end{align}
\begin{align}
    \frac{\partial}{\partial \lambda_{h}} \frac{\partial^2}{\partial x_n \partial x_{n+h}} V 
    = & \frac{1 + \frac{1}{\lambda_{h}^2} + \left(\frac{1}{\lambda_{h}^2} + \frac{3}{\lambda_{h}^4}\right) \log\frac{x_n}{x_{n+h}} - \frac{1}{\lambda_{h}^4}\log^2 \frac{x_n}{x_{n+h}} - \frac{1}{\lambda_{h}^6}\log^3 \frac{x_n}{x_{n+h}} }{ x_n^3 x_{n+h}} \phi (a_{h}) + 
    \notag\\
    & \frac{1 + \frac{1}{\lambda_{h}^2} + \left(\frac{1}{\lambda_{h}^2} + \frac{3}{\lambda_{h}^4}\right) \log\frac{x_{n+h}}{x_n} - \frac{1}{\lambda_{h}^4}\log^2 \frac{x_{n+h}}{x_n} - \frac{1}{\lambda_{h}^6}\log^3 \frac{x_{n+h}}{x_n}  }{x_n^3 x_{n+h}} \phi (a_{h'})
\end{align}

The third factor in the contribution of a single point to (\ref{eq:score_contrib_ap}) is model dependent. We give the forms for each of our motivating models here. The TL-AR(1) has one parameter typically denoted $\phi$ and has TPD function (\ref{eq:AR1_TPD}) which has derivative 
\begin{align}
    \frac{\partial}{\partial \theta_1} \sigma(h, \theta_1) = & 
    \frac{\partial}{\partial \phi}  \phi^h 
    = h\phi^{h-1}. 
\end{align}
The TL-MA(q) has $q$ parameters $(\theta_1, \dots, \theta_q)$ and TPD given by (\ref{eq:MAq_TPD}) which has derivative
\begin{align}
    \frac{\partial}{\partial \theta_k} \sigma(h, \theta_1, \dots, \theta_q) = & \frac{\theta_{k+h}^{(0)} + \theta_{k-h}^{(0)}}{\sum_{l = 0}^q \theta_l^2} + \frac{2\theta_k \sum_{l = 0}^q \theta_l^{(0)}\theta_{l+h}^{(0)}}{\left(\sum_{l = 0}^q \theta_l^2\right)^2}
\end{align}
where one or both of $\theta_{k+h}^{(0)}$ and $\theta_{k-h}^{(0)}$ may be zero. The TL-ARMA(1,1) model has two parameters typically denoted $\phi$ for the AR component and $\theta$ for the MA component. The TPD is given by (\ref{eq:ARMA11_TPD}) which has derivative with respect to $\phi$ 
\begin{align}
    \frac{\partial}{\partial \phi}\sigma(h, \theta, \phi) = &
    \begin{cases}
        \frac{(h - 1)\phi^{h-2}(\phi + \theta)(\phi \theta + 1) + \phi^{h-1}\theta(\phi + \theta) + \phi^{h-1}(\phi \theta + 1)}{\theta^2 + 2\phi \theta + 1} - 
        \notag\\
        \hspace{0.2in} \frac{2 \phi^{h-1} \theta (\phi + \theta)(\phi \theta + 1)}{( \theta^2 + 2\phi \theta + 1)^2} & \text{if } \phi > 0, \phi + \theta > 0
        \notag\\
        0 & \text{if } \phi > 0, \phi + \theta < 0
        \notag\\
        \frac{ h \phi^{h-1}(\phi + \theta)^2 + 2\phi^h(\phi + \theta)}{(\phi + \theta)^2 - \phi^4 + 1} -
        \notag\\
        \hspace{0.2in} \frac{\phi^h (\phi + \theta)^2 (2\phi + 2\theta - 4 \phi^3)}{ \{(\phi + \theta)^2 - \phi^4 +  1\}^2 } & \text{if } \phi < 0, \phi + \theta > 0, h \text{ is even}
        \notag\\
        \frac{(h - 1) \phi^{h - 2} (\phi + \theta) (1 - \phi^{4}) + \phi^{h - 1} (1 - \phi^{4}) - 4\phi^{h + 2} (\phi + \theta)}{(\phi + \theta)^{2} - \phi^{4} + 1}  -  
        \notag\\
        \hspace{0.2in} \frac{\phi^{h - 1} (\phi + \theta) (1 - \phi^{4}) (2\phi + 2\theta - 4\phi^{3})}{\{(\phi + \theta)^{2} - \phi^{4} + 1\}^{2}} & \text{if } \phi < 0, \phi + \theta > 0, h \text{ is odd}
        \notag\\
        \frac{(h - 1) \phi^{h - 2} (\phi + \theta) (\phi^{3}\theta + 1) + \phi^{h - 1} (\phi^{3}\theta + 1) + 3\phi^{h + 1} \theta (\phi + \theta)}{\phi^{2}\theta^{2}  + 2\phi^{3}\theta + 1} - 
        \notag\\ 
        \hspace{0.2in} \frac{\phi^{h - 1} (\phi + \theta) (\phi^{3}\theta + 1) (2\phi\theta^{2} + 6\phi^{2}\theta) }{(\phi^{2}\theta^{2}  + 2\phi^{3}\theta + 1)^{2}} & \text{if } \phi < 0, \phi + \theta < 0, h \text{ is even}
        \notag\\
        0 & \text{if } \phi < 0, \phi + \theta < 0, h \text{ is odd}.
    \end{cases}
\end{align}
The derivative of (\ref{eq:ARMA11_TPD}) with respect to $\theta$ is 

\begin{align}
    \frac{\partial}{\partial \theta} \sigma(h, \theta, \phi) = &
    \begin{cases}
        \frac{\phi^{h - 1} (\phi^2 - 1)(\theta^2 - 1)}{( \theta^2 + 2 \phi \theta + 1)^2} & \text{if } \phi > 0, \phi + \theta > 0
        \notag\\
        0 & \text{if } \phi > 0, \phi + \theta < 0
        \notag\\
        \frac{2 \phi^h (\phi + \theta)}{ (\phi + \theta)^2 - \phi^4 + 1} - \frac{2 \phi^h (\phi + \theta)^3}{( (\phi + \theta)^2 - \phi^4 + 1)^2} & \text{if } \phi < 0, \phi + \theta > 0, h \text{ is even}
        \notag\\
        \frac{\phi^{h-1}(1 - \phi^4)}{(\phi + \theta)^2 - \phi^4 + 1} - \frac{2 \phi^{h-1} (\phi + \theta)^2(1 - \phi^4)}{\{(\phi + \theta)^2 - \phi^4 + 1\}^2} & \text{if } \phi < 0, \phi + \theta > 0, h \text{ is odd}
        \notag\\
        \frac{\phi^{h - 1} (\phi^{4} - 1) (\phi^{2} \theta^{2} - 1)}{(\phi^{2} \theta^{2} + 2\phi^{3} \theta + 1)^{2}} & \text{if } \phi < 0, \phi + \theta < 0, h \text{ is even}
        \notag\\
        0 & \text{if } \phi < 0, \phi + \theta < 0, h \text{ is odd}.
    \end{cases}
\end{align}

\newpage 

\section{Appendix 4: Model Selection}\label{ap:model_selection}

The proxy-likelihood provides us with an objective function which can be used with standard model selection techniques like cross-validation and AIC. 
We caution the reader that the assumptions behind AIC are violated and thus standard properties do not hold. 
Despite this, we have found the penalty to be useful in simulations (Section~\ref{sec:proxy_model_selection}). 
A more principled approach acknowledges that our composite likelihood approach treats each bivariate margin as if it were independent of all other margins and thus it reuses data. 
This results in a surface which, likely, has more curvature than the true likelihood surface suggesting more confidence in parameter estimates than indicated by the data. 
We correct for this by penalizing the proxy-likelihood with the composite likelihood penalty (see, e.g., \citealp{varin2008composite}) which can be thought of as an effective number of parameters.
This penalty is the sandwich variance estimator of \citet{godambe1960optimum} and is used in estimating equations.

We briefly review the estimating equations approach to parameter estimation here in an attempt to highlight the assumptions that we are making in this section. 
For a more thorough discussion see, e.g., \citealp{boos_stefanski2013essential}. 
The estimating equation approach says that if one can come up with a function $\psi({\bf x}_i, \btheta)$ for a single point ${\bf x}_i$ such that the unique solution to $E[\psi({\bf x}_i, \btheta)] = {\bf 0}$ is the parameter of the distribution $\btheta_0$, then we can use $\psi$ to develop an estimator $\hat{\btheta}$ which is a function of the mean of $\psi$ evaluated over several points such that, by the weak law of large numbers, $\hat{\btheta} \overset{p}{\rightarrow} \btheta_0$. 
Estimators developed this way are called M-estimators. 
Furthermore, standard Taylor expansion arguments for $\frac{1}{n} \sum_{i = 1}^n \psi({\bf x}_i, \btheta)$ around $\btheta_0$ demonstrate the asymptotic normality of $\sqrt{n} (\hat{\btheta} - \btheta_0)$ with variance ${\bf H}(\btheta_0)^{-1} {\bf J}(\btheta_0) \{{\bf H}(\btheta_0)^{-1}\}^T$ where ${\bf H(\btheta_0)} = E[- \psi'({\bf x}_i, \btheta_0)]$ and ${\bf J}(\btheta_0) = E[\psi({\bf x}_i, \btheta_0)\psi({\bf x}_i, \btheta_0)^T]$. 

We naturally consider the derivative of the score function (\ref{eq:biv_cllhood}) as $\psi$ and thus ${\bf H(\btheta)}$ is the Hessian and ${\bf J}(\btheta)$ is the covariance of the score function. 
Our context has a few challenges which must be acknowledged. 

First, we do not know the data-generating likelihood; our method is designed to be useful as it captures the second order tail dependence without knowledge of the likelihood. 
For this reason we cannot show that the unique solution to the expected value of our $\psi$ is $\btheta_0$. 
As such we do not claim to have developed an M-estimator but we think that it is illuminating to think of our method in that framework and that the correction to the variance estimator may be useful for model selection. 

Second, (\ref{eq:biv_cllhood}) includes the entire observed time series and thus we do not have replications. 
Without replications we cannot average to get the WLLN result. 
The consistency of the estimator relies on a consistent estimator for 
\begin{align*}
    \mathbf{J}(\hat{\boldsymbol\theta}_{MPL}) = \text{Var}\big[ S(\hat{\boldsymbol\theta}_{MPL}, {\bf X}) \big] 
    = \text{Var}\Bigg[ \sum_{h = 1}^{h_{max}} \sum_{n = 1}^{N-h} S_h(\hat{\boldsymbol\theta}_{MPL}, X_n, X_{n+h}) \Bigg]
\end{align*}
but this is challenging due to the lack of replicates. 
We cannot pull a sum out of the variance because neither sum is over independent random variables. 
We follow \citet{heagerty_lumley2000window} who suggest using the entire time series in the estimator $\hat{\btheta}$ but then using subsets of the domain $1, \dots, N$ to compute the score. 
This allows us to estimate ${\bf J}(\btheta)$ as the covariance of the estimated scores computed over each sub-domain. 
Let $m = 1, \dots, M$ index the subseries of length $n_m$ which we treat as independent replicates. 
With these quasi-replicates we consider
\begin{align*}
    {\bf J}(\hat{\boldsymbol\theta}_{MPL}) \approx& \sum_{m = 1}^M \text{Var}\Bigg[ \sum_{h = 1}^{h_{max}} \sum_{n = 1}^{n_m-h} S_h(\hat{\boldsymbol\theta}_{MPL}, X_{m,n}, X_{m,n+h}) \Bigg]
\end{align*}
which allows us to use the natural estimator (the empirical variance over the subseries). 

Standard arguments for asymptotic normality would rely on a CLT for $\alpha$-mixing processes as our process is clearly not $iid$. 
The $M$ sub-series should satisfy $\alpha$-mixing requirements but this is hardly important because any claims of approximate normality are subject to the prior caveats. 

In practice we split our data into a small number of subseries ($M \in 10, \cdots, 20$) as we need to have tail dependence information in each.
Let $\hat{\btheta}_{MPL} = (\hat\theta_{1, MPL}, \dots, \hat\theta_{K, MPL})$ be the maximum proxy-likelihood (MPL) estimator for $\btheta$. 
The composite likelihood versions of the AIC \citep{varin_vidoni2005note} and BIC \citep{gao_song2010composite} are the natural tool for comparison across models:
\begin{align}
    \text{CLAIC} =& -2\ell_{cl}(\hat{\btheta}_{MPL}|\mathbf{x}) + 2 \text{tr}\{\mathbf{J}(\hat{\btheta}_{MPL}) \mathbf{H}^{-1} (\hat{\btheta}_{MPL})\} \label{eq:claic}\\
    \text{CLBIC} =& -2\ell_{cl}(\hat{\btheta}_{MPL}|\mathbf{x}) + \log(n) \text{tr} \{\mathbf{J}(\hat{\btheta}_{MPL}) \mathbf{H}^{-1} (\hat{\btheta}_{MPL})\}.
\end{align}

The composite likelihood model selection techniques require us to estimate the Hessian ($\mathbf{H}$) and the variance of the score ($\mathbf{J}$). The score equation in Appendix~\ref{ap:score_func} allows for natural covariance estimation. We could write down an estimator of the Hessian through differentiating the score function but it is easily obtained from many standardized computer optimization routines which is what we use in practice. 

While having straightforward uncertainty estimation through confidence intervals is always desirable, and is a common motivator for likelihood methods, we do not prioritize uncertainty quantification. 
One could use the estimated matrices $\hat{\mathbf{J}}(\hat{\btheta}_{MPL})$ and $\hat{\mathbf{H}}(\hat{\btheta}_{MPL})$ in a sandwich variance estimator.
Standard arguments could then be used to develop confidence intervals for the parameters in the models that we fit. 
Initial investigation suggests that coverage of intervals made using these standard arguments is poor. 
This is not a big concern for us because our primary concern is the second-order dependence property (the TPD) not the parameters. 
If uncertainty quantification is desired we suggest using bootstrapping.

\newpage 

\section{Appendix 5: Censoring}\label{ap:censoring}
\citet{huser_davison_genton_2016} show that using Euclidean threshold censoring in pairwise likelihood estimation has lower bias and RMSE than other proposed likelihood estimators. In this section, we develop this censoring scheme in our context.  Consider the data vector $(X_1, \cdots, X_n)$ which, after transformation, is marginally Frechet(2). Let $u_q$ be the $q$-quantile of a Frechet(2) distribution which will be our censoring threshold. Following \citet{smith_tawn_coles1997markov} we let $\delta_i = \mathbb{I}(X_i > u_q)$ and $Y_i = \max(0, X_i - u_q)$. The joint distribution of $(\delta_i, Y_i, \delta_j, Y_j)$ is

\begin{align}\label{eq:euc_censored_lhood}
    g_{\lambda_{ij}}^c(\delta_i, y_i, \delta_j, y_j) 
    = &
    \begin{cases}
       G_{\lambda_{ij}} (u_q, u_q) \hspace{0.1in} & \delta_i = \delta_j = 0
       \\
       \frac{\partial}{\partial x_i} G_{\lambda_{ij}} (u_q + y_i, u_q) \hspace{0.1in} & \delta_i = 1, \delta_j = 0
       \\
       \frac{\partial}{\partial x_j} G_{\lambda_{ij}} (u_q, u_q + y_j) \hspace{0.1in} & \delta_i = 0, \delta_j = 1
       \\
       \frac{\partial^2}{\partial x_i \partial x_j}G_{\lambda_{ij}} (u_q + y_i, u_q + y_j) \hspace{0.1in} & \delta_i = \delta_j = 1.
    \end{cases}
    \\
    = &
     \begin{cases}
        G_{\lambda_{ij}} (u_q, u_q) \hspace{0.1in} & \delta_i = \delta_j = 0
        \\
        \bigg[\frac{2}{(u_q + y_i)^3} \Phi (a_{ij}) + \frac{1}{\lambda_{ij} (u_q + y_i)^3} \phi (a_{ij}) \\
        \hspace{0.5in} - \frac{1}{\lambda_{ij} (u_q + y_i) u_q^2} \phi (a_{ji})\bigg] G_{\lambda_{ij}} (u_q + y_i, u_q) \hspace{0.1in} & \delta_i = 1, \delta_j = 0
        \\
        \bigg[\frac{2}{(u_q + y_j)^3} \Phi (a_{ji}) + \frac{1}{\lambda_{ij} (u_q + y_j)^3} \phi (a_{ji}) \\
        \hspace{0.5in} - \frac{1}{\lambda_{ij} (u_q + y_j) u_q^2} \phi (a_{ij})\bigg] G_{\lambda_{ij}} (u_q, u_q + y_j) \hspace{0.1in} & \delta_i = 0, \delta_j = 1
        \\
        g_{\lambda_{ij}} (u_q + y_i, u_q + y_j) \hspace{0.1in} & \delta_i = \delta_j = 1.
    \end{cases}
\end{align}
Where the derivatives in (\ref{eq:euc_censored_lhood}) are in (\ref{eq:density_derivs_1_app}) and $\partial^2 G_{\lambda_{ij}}/ \partial x_i \partial x_j (\cdot) = g_{\lambda_{ij}} (\cdot)$ (from (\ref{eq:HR_density_lambdas})). 

This censored likelihood requires a new first component of the score function (\ref{eq:score_contrib_ap}). While this censoring approach has four cases (both dimensions are small, the $i^{th}$ component is small, the $j^{th}$ component is small, or both dimensions are large) we only have two new cases to derive. The two one-small cases are the same by symmetry and the both large case is the same as (\ref{eq:component_one}). The case when both dimensions are small is 

\begin{equation}
    \frac{\partial}{\partial \lambda} \ell^c_{cl}(\lambda | \delta_i = \delta_j = 0) = - \frac{\partial}{\partial \lambda} V(u_q, u_q) = -2u_q^{-2} \phi(\lambda).
\end{equation}

When one component is small (WLOG we assume $\delta_i = 1$) the score contribution is 

\begin{align}
    \frac{\partial}{\partial \lambda} & \ell^c_{cl}(\lambda | \delta_i = 1, \delta_j = 0) 
    \notag\\ 
    = &  \frac{\partial}{\partial \lambda} \log\bigg[ \frac{2}{(u_q + y_i)^3} \Phi (a_{ij}) + \frac{1}{\lambda_{ij} (u_q + y_i)^3} \phi (a_{ij})- \frac{1}{\lambda_{ij} (u_q + y_i) u_q^2} \phi (a_{ji}) \bigg]
    \notag\\ 
    = & \bigg[\frac{2}{(u_q + y_i)^3} \Phi (a_{ij}) + \frac{1}{\lambda_{ij} (u_q + y_i)^3} \phi (a_{ij})- \frac{1}{\lambda_{ij} (u_q + y_i) u_q^2} \phi (a_{ji})\bigg]^{-1} 
    \notag\\
    & \hspace{0.1 in} * \big\{[1 - \lambda^{-2} + (2\lambda^{-2} + \lambda^{-4}\log(x_i/u_q))\log(x_i/u_q)]  x_i^{-3} \phi(a_{ij})
    \notag\\
    & \hspace{1.2in} + [\lambda^{-1} + \lambda - \lambda^{-3} \log^2(u_q/x_i)] x_i^{-1} u_q^{-2}\phi(a_{ji})]\big\}
\end{align}

\end{document}